# A review on substellar objects beyond the deuterium burning mass limit: planets, brown dwarfs or what?

José A. Caballero [1]

[1] Centro de Astrobiología (CSIC-INTA), ESAC, Camino Bajo del Castillo s/n E-28692 Villanueva de la Cañada, Madrid, Spain; caballero@cab.inta-csic.es

**Abstract:** "Free-floating, non-deuterium-burning, substellar objects" are isolated bodies of a few Jupiter masses found in very young open clusters and associations, nearby young moving groups and in the immediate vicinity of the Sun. They are neither brown dwarfs nor planets. I look over their nomenclature, history of discovery, sites of detection, formation mechanisms and future directions of research. Most free-floating, non-deuterium-burning, substellar objects share the same formation mechanism as low-mass stars and brown dwarfs, but there are still a few caveats, such as the value of the opacity mass limit, the minimum mass at which an isolated body can form via turbulent fragmentation from a cloud. The least massive free-floating substellar objects found to date have masses of about 0.004 $M_{sol}$, but current and future surveys should aim at breaking this record. For that, we may need LSST, *Euclid* and *WFIRST*.

## 1. Introduction

*I can't answer why (I'm not a gangstar)*
*But I can tell you how (I'm not a flam star)*
*We were born upside-down (I'm a star's star)*
*Born the wrong way 'round (I'm not a white star)*
*I'm a blackstar, I'm not a gangstar*
*I'm a blackstar, I'm a blackstar*
*I'm not a pornstar, I'm not a wandering star*
*I'm a blackstar, I'm a blackstar*

Blackstar, * (2016), David Bowie

The tenth star of the George van Biesbroeck's catalogue of high, common, proper motion companions, vB 10, was from the end of the Second World War to the early 80s the least massive star known [vB44, Her56, Kir91]. At only 6 pc (actually 5.91 pc with recent *Gaia* DR2 data [Gai18]) and with an M8.0 V spectral type [AF15, Kam18], the mass of vB 10 was estimated at about 0.08 $M_{sol}$, at the theoretical limit for sustaining nuclear fusion of ordinary hydrogen ($^1$H). The red dwarf had not only a very low mass, but it was also very cold ($T_{eff}$ ~ 2600 K) and very small ($R$ ~ 0.11 $R_{sol}$) and, therefore, had a very low luminosity ($L$ ~ 0.0004 $L_{sol}$). Although later its mass was revised to above 0.09 $M_{sol}$ [Sch18a] and LHS 2924, first [PL83], and many other red dwarfs, afterwards, took its place as "the least massive star in the solar neighborhood", the existence of vB 10 challenged both observational and theoretical astrophysicists for four decades.

Is there a mass limit for the least luminous stars? Is there a bridge between the least massive stars and the most massive planets? Are there planets larger than Jupiter? How is the physics inside such bodies? In the early 60s these questions fluttered in the heads of several theoretical astrophysicists [Kum03]. In a pioneer work published in 1963, Shiv S. Kumar constructed a grid of completely convective models of "stars" of masses from 0.09 $M_{sol}$ to 0.04 $M_{sol}$ and showed that there is a lower limit to the mass of the stellar main sequence. In his words:

*"The stars with mass less than this limit become completely degenerate stars or 'black' dwarfs as a consequence of gravitational contraction, and, therefore, they never go through the normal stellar evolution"* [Kum63].

To this, Chushiro Hayashi and Takenori Nakano added in the same year that:

*"The stars less massive than* 0.08 $M_{sol}$ *are found to contract toward the configurations* [sic] *of high electron-degeneracy without hydrogen burning"* [HK63].

That is, Kumar, Hayashi and Nakano predicted that there could be substellar objects beyond the bottom of the stellar main sequence but, if they existed, they would be degenerate, completely convective, not able to burn hydrogen in their interiors, and would cool down and get faint for ever. To avoid the confusion with actual "black dwarfs", i.e. theoretical extremely old white dwarfs that will cool sufficiently that their radiation in the visible and near-infrared will hardly be detected [MR67, Vil71], in 1975 Jill Tarter coined the name "brown dwarfs" for such substellar objects [Tar14].

The publication of less and less massive stars and some controvertible brown dwarf candidates populated astronomy journals in the late 80s and early 90s: PC 0025+0047 and DY Psc (BRI 0021-0214), dwarfs with spectral spectral type "M9 or later" [Irw91, Sch91]; GD 165 B, recognized as an L dwarf companion to a white dwarf over a decade after its discovery [BZ88, Kir99a]; HD 114762 b, a radial-velocity companion to an F9 V star with a minimum mass of only 0.011 $M_{sol}$ [Lat89]; vB 8 B and ZZ Psc B, fake brown dwarfs based on poor quality data [McCP84, ZB87]; LP 944-20, a very late field M dwarf discovered in 1975 that could be very young and, thus, have a substellar mass [LK75, Tin98]; [RR90] Oph 2321.1-1715, 2408.6-2229 and 2412.9-2447, three reddened, embedded sources in the core of ρ Ophiuchi and whose nature is still debated [RR90, Com98], and many more.

In 1994 *"there was a palpable sense of frustration at the failure of many efforts to confirm a single brown dwarf"* [Bas14]. However, in the following year, the same miraculous year when the first exoplanet was found in orbit to a main sequence star, 51 Peg b [MQ95], the first two uncontrovertible brown dwarfs appeared on stage: Teide 1, a free-floating high-mass brown dwarf that passed the lithium test in the young Pleiades cluster [Reb95, Reb96] (published just before PPl 15 AB, a pair of true brown dwarfs also in the Pleiades [Bas96]), and GJ 229 B, a T-type wide companion to the nearby star HD 42581 (also known as GJ 229) [Nak95, Opp95].

Since then, we have found hundreds, if not thousands, of substellar objects: from accreting and disk-bearing brown dwarfs in the young σ Orionis open cluster [Béj99, ZO02a, Cab07a], through Luhman 16 AB, a brown dwarf binary at the L-T transition and the third-closest-known system to the Sun after α Centauri and Barnard's star [Luh13], to the numerous ultracool dwarfs in the field with parallaxes measured by the ESA *Gaia* mission [Sma17]. Almost a quarter-century after the discovery and correct identification of the first brown dwarfs, we have compared our theoretical models [Bur97, Bar98, Cha00, All01, Bat03, Bar15, MR15] with our observations [Del99, Bur00, Giz00, BJM01, Bou03, WB03, ZO05, Mor08, Art09, Rad12, DK13], defined three new spectral subtypes for classifying them, L, T and Y [Kir99b, Bur02, Geb02, Kir05, Kir12], and found that some substellar objects can have effective temperatures as cool as 225-260 K (e.g. WISE 085510.83-071442.5 [Luh14a])… A temperature colder than a Siberian winter night!

By definition, brown dwarfs do not burn the common isotope of hydrogen ($^1$H) through the proton-proton chain reaction. However, all of them do burn deuterium ($^2$H) and, only the most massive ones, lithium (especially the most abundant isotope, $^7$Li). Similarly to main-sequence stars, the lower the mass of the brown dwarf, the lower its central temperature. The minimum temperature for hydrogen burning is about $3 \cdot 10^6$ K, which translates at solar metallicity into a theoretical substellar boundary mass of 0.072 $M_{sol}$. It is the hydrogen burning mass limit or, in other words, the bottom of the main sequence. Lithium nuclei are destroyed by collisions with protons at slightly lower central temperatures of about $2 \cdot 10^6$ K. According to the Lyon theoretical models [Cha00'], a 0.06 $M_{sol}$-mass object of solar metallicity and age destroys all its lithium, but a 0.05 $M_{sol}$-mass one preserves 92% of the original lithium content (this is the basis of the lithium test for assessing the substellar nature of an object: after a certain time, if its spectrum displays the Li I $\lambda$6707.8 Å resonant doublet line –and there has not been lithium dragging or other exotic process–, then the object is indeed a brown dwarf [Reb92, Mag93]).

The minimum temperature for deuterium burning is much lower, of only $0.5 \cdot 10^6$ K (while the proton-proton chain is driven by the weak nuclear force, the primary reaction of the deuterium thermonuclear fusion, $D(p+,\gamma)^3He$, is driven by the electromagnetic force [Sau96, Spi11]). A brown dwarf of 0.013 $M_{sol}$ and solar metallicity can reach, and only for a brief duration at the early stages of evolution, such an internal temperature. Any substellar object below that mass, the deuterium burning mass limit, will not be able to sustain any nuclear fusion reaction in their interior (Fig. 3).

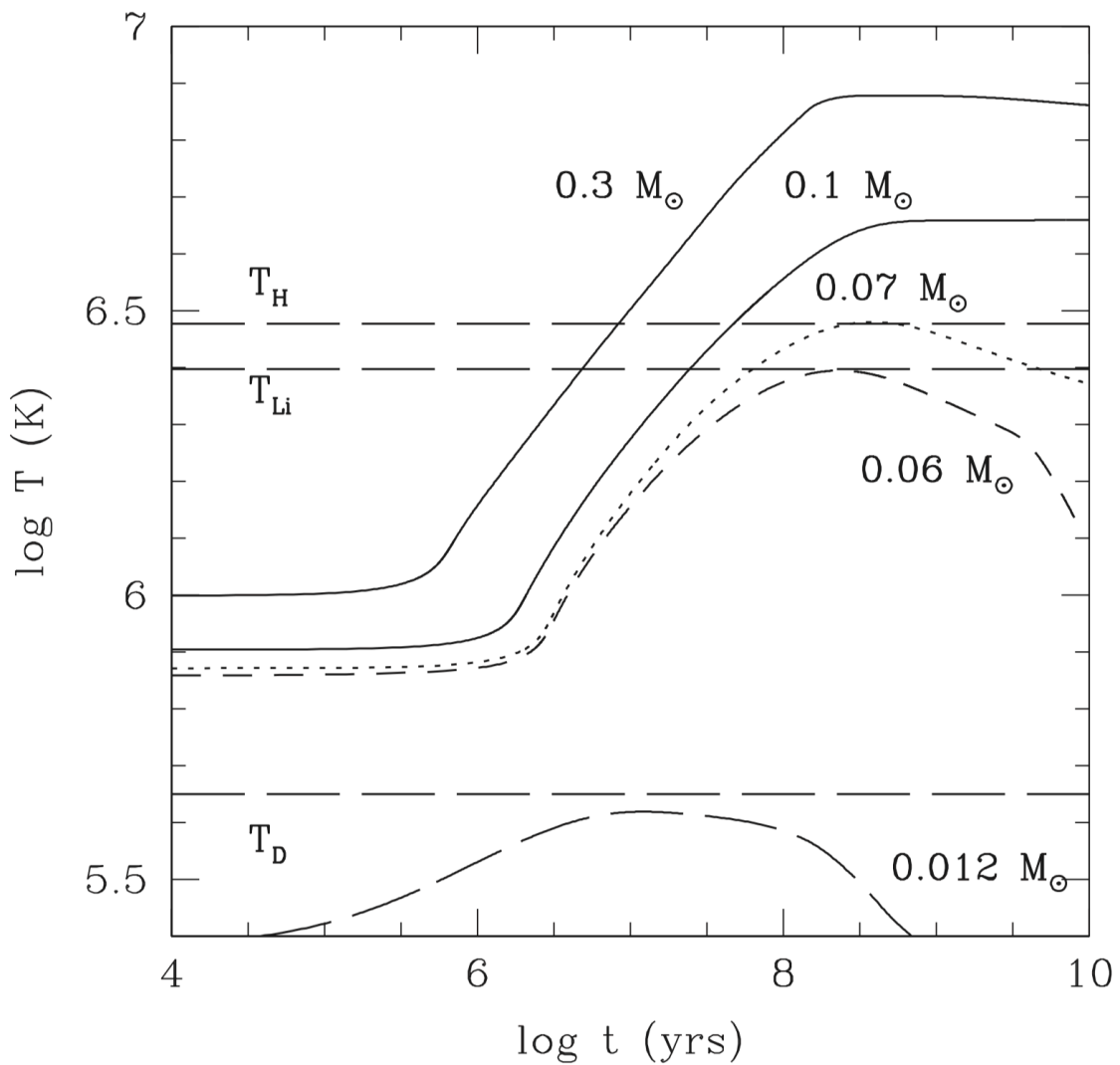

**Figure 1.** Central temperature of stars, brown dwarfs and "planetary-mass objects" as a function of age for 0.3, 0.1, 0.07, 0.06 and 0.012 $M_{sol}$, from top to bottom. Dashed horizontal lines mark the burning temperatures for $^1H$ hydrogen, $^7Li$ lithium and $^2H$ deuterium. "Planetary-mass objects" never reach a central temperature for deuterium fusion. Reproduced with permission from [CB00].

In parallel to the brown dwarf discoveries of the last two decades and a half, the exoplanet hunters have competed against themselves in the search for the radial-velocity planet with the lowest minimum mass, the transiting planet with the smallest radius, the planet with the most suitable atmosphere for characterization, the planet with the most similar conditions to the Earth, the closest planetary system, the planetary system with most planets [Cha00, Cha02, But04, Bea06, Knu07, Udr07, Bat11, Doy11, Qui14, Mac15, AE16, Gil17]… In a 21st century version of the Olympic motto *Citius, altius, fortius*, exoplanet hunters pursue not only "more", but "the most". And in this pursuit they also find "superjupiters": exoplanets with masses several times greater than that of Jupiter. These superjupiters are found around stars in close orbits (with periods of a few days), mostly with spectroscopy and the radial velocity method, and in wide orbits (with periods of decades and centuries), mostly with direct imaging and adaptive optics. Examples of such superjupiters are, on one side, 70 Vir b and τ Boo Ab, with minimum masses of about 0.005-0.007 $M_{sol}$ [MB96, But97], and, on the other side, the four planetary companions of HD 218396 (also known as HR 8799), with masses between 0.004 $M_{sol}$ and 0.010 $M_{sol}$ [Mar08, Mar10]. (Actually, exoplanet hunters use Jupiter masses, $M_{Jup}$, or even Earth masses, $M_{Terra}$, instead of solar masses: 1 $M_{sol}$ = 1047.6 $M_{Jup}$; 1 $M_{Jup}$ = 317.8 $M_{Terra}$).

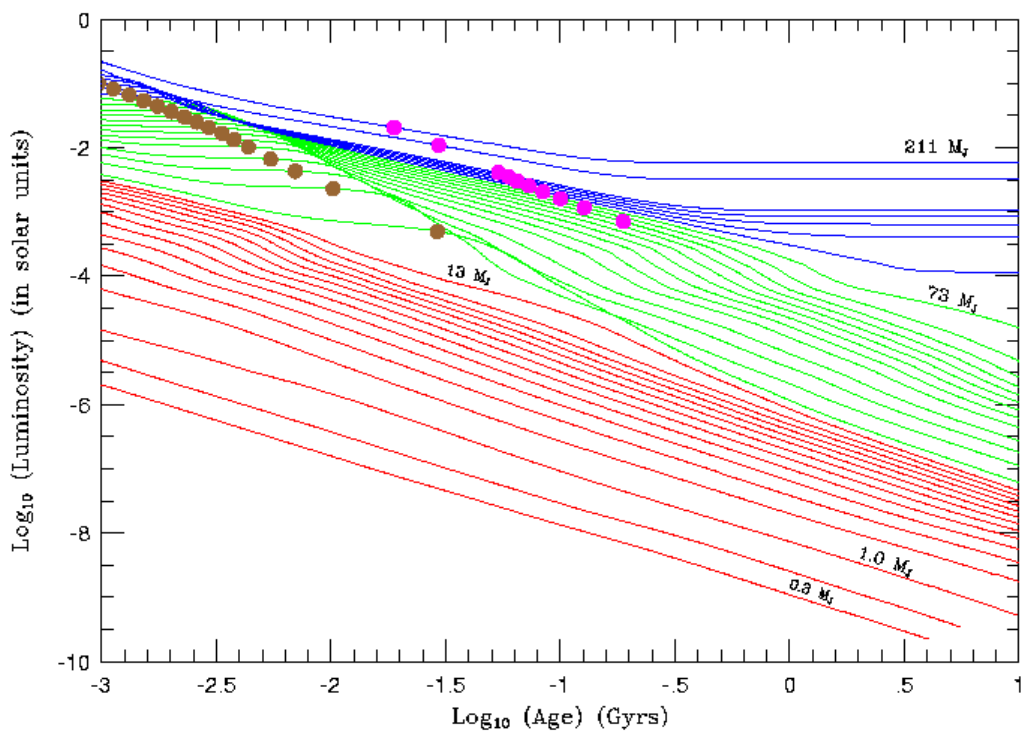

**Figure 2.** Cooling curves for M dwarfs (blue), brown dwarfs (green), and "extrasolar giant planets" (red) according to the evolutionary models of [Bur97']. Magenta and brown circles indicate the times when deuterium fusion ends and grain formation begins, respectively. "Planetary-mass objects" of a few megayears can have luminosities greater than low-mass stars of a few gigayears. Heritage scientific graphic downloaded from Adam Burrows' home page [bur].

Now we can interrogate ourselves again: Is there a mass limit for the least luminous *brown dwarfs*? Is it the deuterium burning mass limit? Is there a bridge between the least massive *brown dwarfs* and the most massive planets? How do we call free-floating substellar objects that do not fuse deuterium and have masses similar to those of superjupiters? Are they planets or brown dwarfs? Where and how do we find them? How do they form? How do they evolve? How is the physics inside such bodies? Here I try to answer these questions and review the most important ideas on free-floating, non-deuterium-burning, substellar objects detected with direct imaging (see [Tsa18] in this volume for a review on free-floating, non-deuterium-burning, substellar objects detected with microlensing).

## 2. Nomenclature

Finding an appropriate, short name for our "free-floating, non-deuterium-burning, substellar objects" is a dilemma. Here there are some of the names used for designating them:

- Cluster planet
- Directly-imaged gas-giant planet
- Free-floating planet
- Free-floating planetary-mass brown dwarf
- Interstellar planet
- Isolated extrasolar giant planet
- Isolated planetary-mass object
- Nomad planet
- Orphan planet
- Plamo (contraction of "planetary-mass object")
- Planemo (idem)
- Planetar (originally coined for designating brown dwarfs)
- Rogue planet
- Starless planet
- Sub-brown dwarf
- Sunless planet
- Superjupiter
- Wandering planet

In 2003, the Working Group on Extrasolar Planets (WGESP) of the International Astronomical Union presented a position statement on the definition of a planet, which is complementary to the definition voted in Prague in 2006 of a Solar System planet (the one in which Pluto became a dwarf planet). WGESP agreed on the following [Bos07]:

1. *Objects with true masses below the limiting mass for thermonuclear fusion of deuterium (currently calculated to be 13 Jupiter masses [0.013 $M_{sol}$] for objects of solar metallicity) that orbit stars or stellar remnants are "planets" (no matter how they formed). The minimum mass/size required for an extrasolar object to be considered a planet should be the same as that used in our Solar System.*
2. *Substellar objects with true masses above the limiting mass for thermonuclear fusion of deuterium are "brown dwarfs", no matter how they formed nor where they are located.*
3. *Free-floating objects in young star clusters with masses below the limiting mass for thermonuclear fusion of deuterium are not "planets", but are "sub-brown dwarfs" (or whatever name is most appropriate).*

This position statement is illustrated by the flowchart in Fig. 2. As emphasized by WGESP, this was a working definition and "*a compromise between definitions based purely on the deuterium-burning mass or on the formation mechanism*". WGESP also "*expected this definition to evolve as our knowledge improves*" [Bos07']. This definition does not include free-floating, non-deuterium-burning, substellar objects around a brown dwarf (Section 3.2). The first WGESP name proposal for substellar objects that are neither brown dwarfs or planets was "sub-brown dwarf". However, this term may lead to confusion with "brown subdwarfs", i.e. low-metallicity ultracool dwarfs of mid-L spectral type or later that extrapolate the luminosity class VI at the bottom of and slightly below the main sequence [Bur03a, Bur04a, Scho04, Lod17, Zha17a, Zha17b.]. All other names enumerated at the beginning of this section carry the chain "plane-" (e.g. rogue planet, isolated planetary-mass object, planemo or "brown" (e.g. free-floating planetary-mass brown dwarf). Given the difficulty in reading "free-floating, non-deuterium-burning, substellar object" or its corresponding acronym (FFNDBSO), that this review is part of a special issue on "Detection and characterization of extrasolar planets", and for emphasizing that such bodies are not physically bound to any star or stellar remnant, hereafter I will use the term "isolated planetary-mass object" or, better, the acronym iPMO for designating them.

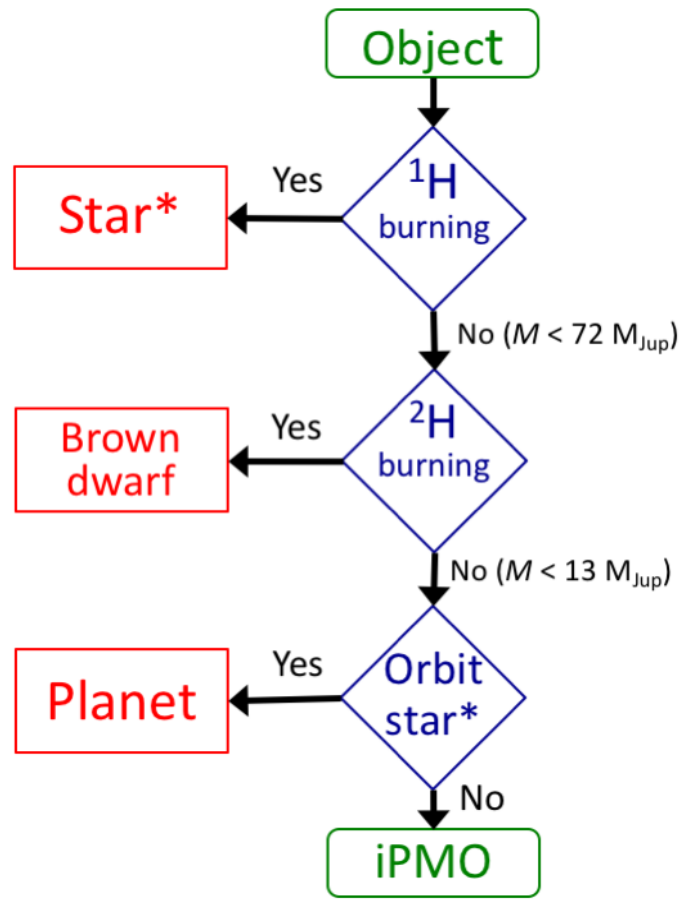

**Figure 3.** Flowchart of the Working Group on Extrasolar Planets position statement on the definition of a planet (1 $M_{sol}$ = 1047.6 $M_{Jup}$). The acronym iPMO stands for "isolated planetary-mass object", but it can be replaced by "*sub-brown dwarf (or whatever name is most appropriate)*". The asterisks of stars indicate that some of them also burn helium and more massive isotopes, and that planets orbit stars, white dwarfs, neutron stars or black holes. Image courtesy of the author.

The planet/brown dwarf/iPMO classification used here and by WGESP is in contradiction with the physical criterion for including a candidate in the most relevant exoplanet catalogue. To date, The Extrasolar Planets Encyclopaedia [exo'] tabulates 3820 exoplanet candidates in 2854 planetary systems. The Encyclopaedia former upper mass limits were 0.013 $M_{sol}$, based on the deuterium burning limit, and 0.030 $M_{sol}$, based on some formation scenarios (Section 5). However, the current limit is 0.060 Msol + 1σ because, as pointed out by Hatzes & Rauer [HR15], "*the mass-density-radius distribution shows a clear difference between giant planets and stars at 60 $M_{Jup}$ [0.060 $M_{sol}$]*". As a result, the Encyclopaedia currently lists many brown dwarfs according to the WGESP definition (either single, as Teide 1 [Reb95'], or companions, as GJ 229 B [Nak95']), as well as a few nearby iPMO candidates, which may also lead to confusion to newcomers (Fig. 3).

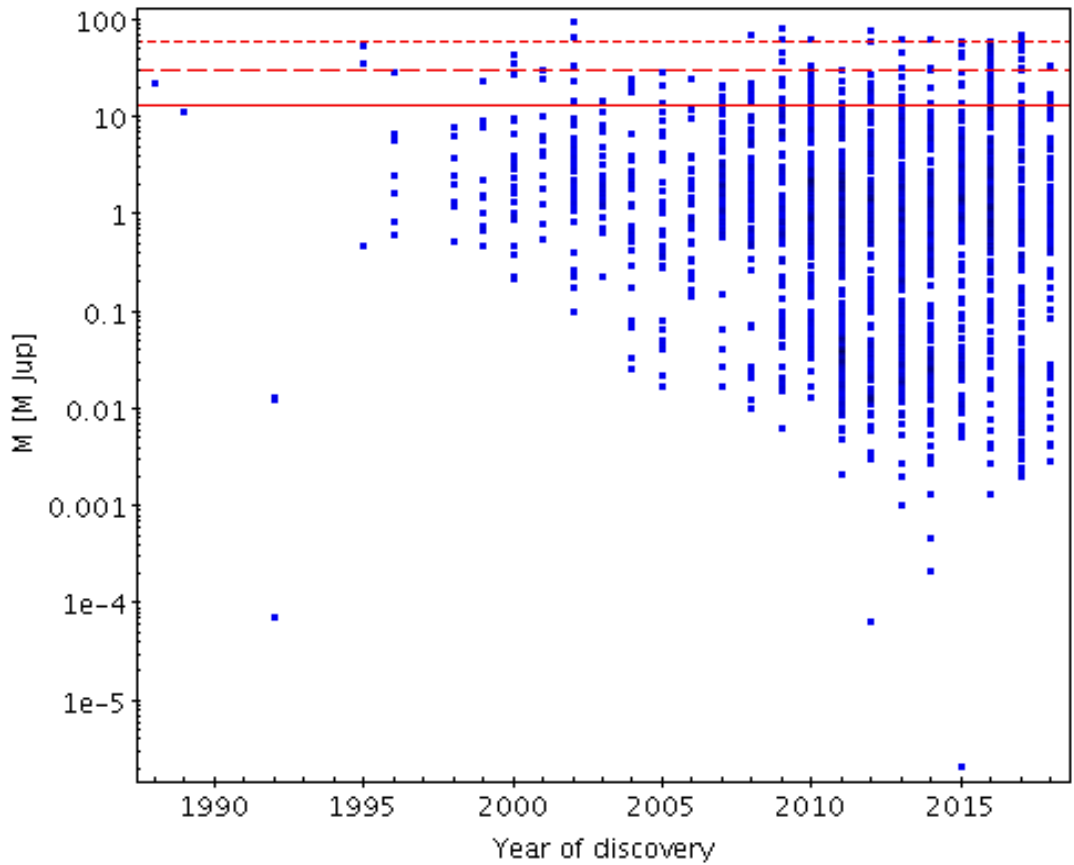

**Figure 4.** Mass or minimum mass of objects catalogued in The Extrasolar Planets Encyclopaedia as a function of year of discovery until August 2018 [exo]. The horizontal lines mark the different Encyclopaedia upper mass limits at 0.060 $M_{sol}$ (dotted), 0.030 $M_{sol}$ (dashed) and 0.013 $M_{sol}$ (solid), from top to bottom. Objects above the solid line (±1σ) are brown dwarfs or stars according to the WGESP definition of planet. Below the solid line there are also a few iPMO candidates. Image courtesy of the author.

## 3. History of discovery

The discovery of the first brown dwarfs was a natural consequence of the development of astronomical instrumentation and techniques and the use of larger telescopes with respect to previous searches of low-mass stars. Likewise, the discovery of the first iPMOs originated from new deep searches for faint brown dwarfs, especially in young open clusters. The deeper the survey, the less massive the cluster members. The main aim of these searches was to determine the shape and end of the initial mass function [Sal55, Sca86, Kro01, Cha03] at very low masses, down to below the hydrogen burning limit.

The first iPMOs were published in 2000 in the Orion Nebula Cluster (ONC) by Lucas & Roche [LR00] and the σ Orionis open cluster by Zapatero Osorio et al. [ZO00], in this order. The UK team did not present any spectroscopic data of their targets until one year later [Luc01], while the Spanish one did it already. Since at least spectral type determination is necessary at these very low masses for identifying possible contaminants, both European teams [LR00', ZO00'] are now widely recognized by their quasi-simultaneous discoveries. One year before, another Japanese team had also found a few embedded targets in the clouds of Chamaeleon [Oas99], but the high extinction prevented astronomers from confirming their nature for years [Luh04], while [Naj00] reached the deuterium burning mass limit in IC 348, but could not go deeper.

The targets found and the techniques and facilities used by both UK and Spanish teams were quite similar. ONC and σ Orionis are more or less at the same heliocentric distance (d ~ 400 pc) [Cab08a, Rei14, Sch16, Kou17, Bri18, Cab18] and galactic latitude, have solar metallicity [McW97,

GH08], are located in the same direction towards the antapex in the Ori OB1 association, and project the same angular size on the sky of about 1 deg. ONC in the Orion Sword is slightly younger (t ~ 1 Ma) than σ Orionis in the Orion Belt (t ~ 3 Ma) [ZO02a', Pal05, Her07, She08, Jef11]. As a result, the extinction is larger in ONC than in σ Orionis: while one could see all Milky Way stars in the line of sight of the Orion Belt with an ultradeep survey [Cab08b], there is only contamination by foreground sources towards the Orion Sword. The age and extinction effects compensate each other, and the substellar boundary in ONC and σ Orionis roughly lies at the same magnitude *I* =17-18 mag [Cab07a'].

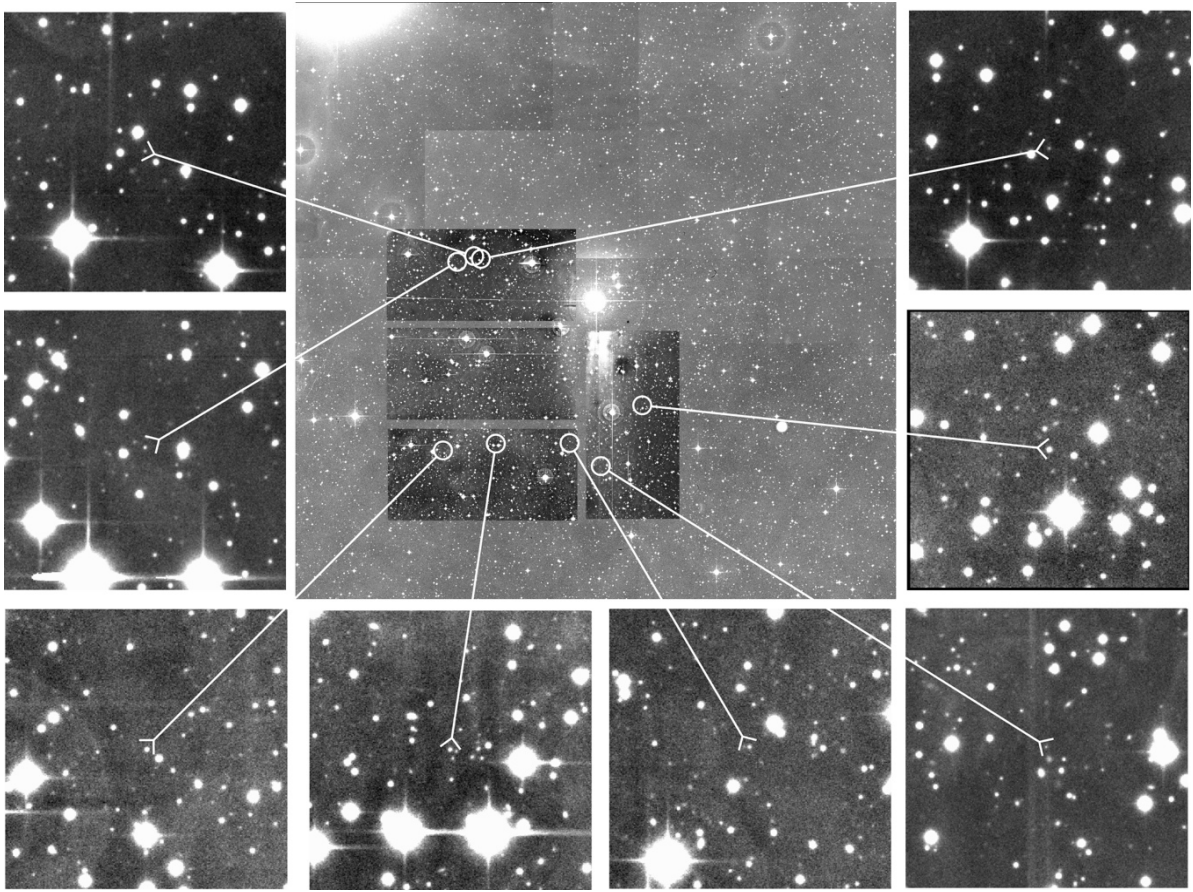

**Figure 5.** Mosaics of *I*-band images of σ Orionis taken with the Wide Field Camera on the 2.5 Isaac Newton Telescope showing the location of some iPMO candidates in [Cab07']. The size of each small inset picture is $2 \times 2$ arcmin$^2$; the size of each Wide Field Camera frame is about $11 \times 22$ arcmin$^2$; the size of the background image from Digitized Sky Survey II Infrared is $1 \times 1$ deg$^2$. Image courtesy of the author.

The two teams used wide-field cameras at 2-4 m telescopes and the reddest optical bands suitable at that moment (*RIZ*). They minimized the number of pointings by surveying the central parts of both clusters, but avoiding the bright Trapezium and Trapezium-like stars at the very centers, which saturate the detectors and generate wide bleeding and smearing lines and optical ghosts (see example of a representative deep survey in Fig. 5). After bias and flat-field corrections, they performed the corresponding photometric analysis on the CCD images (generally, aperture photometry first, and point spread function photometry, next), discarded non-stellar sources, plotted colour-magnitude diagrams, extrapolated the cluster sequence with the help of theoretical isochrones and bona fide cluster members of higher mass, and applied a certain selection criterion in the diagram(s). After compiling a list of brown dwarf and iPMO candidates, the most promising ones were subject of follow-up near-infrared imaging and spectroscopy at larger telescopes. Given the faintness of iPMOs (and extinction in the case of ONC), with red optical magnitudes of *I* = 22-25 mag, spectroscopy could be performed only at 4 m to 10 m-class telescopes, and with low spectral resolution.

Before 2006, very few iPMOs had published spectroscopy because of its faintness. The targets listed in Table 1 consumed hours and hours of research: preparing telescope time proposals, being awarded by time allocation committees, observing in imaging mode in the optical for several consecutive nights (Orion is a winter constellation in the Northern hemisphere, so the fraction of time lost due to bad weather is high), reducing and analyzing the data, selecting iPMO candidates, back to proposals and committees after one semester, back to telescope for near-infrared imaging and spectroscopy, back to reduction and analysis… While today this process can be routine, in the late 90s and early 00s the available hardware (hard disks of more than 8 GB were rare) and software (Python 3.0 was released in 2008) made the process to be slow and poorly automated.

**Table 1.** iPMO candidates with published spectroscopy before 2006 [1].

| Name | $M$ [$M_{Jup}$] | Reference |
|---|---|---|
| S Ori 53 | $14^{+6}_{-7}$ | [ZO00', Cab07a'] |
| S Ori 55 | $12^{+4}_{-4}$ | [ZO00', ZO02b] |
| 61-401 | 12 | [Luc01'] |
| S Ori 56 | 10: | [ZO00'] |
| S Ori 58 | 10: | [ZO00'] |
| S Ori 60 | $8^{+7}_{-3}$ | [ZO00', Cab07a'] |
| Cha 110913-773444 | (8) | [Luh05] |
| 23-115 | 8 | [Luc01'] |
| S Ori 62 | $7^{+7}_{-3}$ | [ZO00', Cab07a'] |
| S Ori 65 | 6 | [ZO00'] |
| S Ori 67 | 6 | [ZO00'] |
| S Ori 68 | 5 | [ZO00'] |
| S Ori 69 | (5) | [ZO00'] |
| 2M1207-29b | $5^{+2}_{-2}$ | [Cha04, Cha05] |
| S Ori 70 | $(3^{+5}_{-1})$ | [ZO02c, MZO03] |

[1] Based on a Table by [Cab06a] (see summary in English in [Cab10]) after deleting star companions GQ Lub b and AB Pic b. Masses have been revised with new data.

Not all iPMO candidates listed in Table 1 are actually free-floating, non-deuterium-burning, substellar objects (with masses in parenthesis). Most of them were found in σ Orionis (they were baptized with the misleading name "S Ori"), but a few of them were also found in ONC, Chamaeleon and as a common proper-motion companion of a young brown dwarf in the TW Hydrae association (Section 3.4). With the extensive use of larger telescopes and deeper surveys, the number of iPMOs with spectroscopy has increased, although not dramatically. However, as we will see below, there is a recently new site for discovering iPMOs, very different from young open clusters and nearby associations, and that has started to bear fruits: just in our immediate vicinity!

## 3. iPMOs here and there

### 3.1. iPMOs in young open clusters

Surveys for iPMOs in open clusters offer advantages over those in the field. All bona fide cluster members, under several assumptions, share a common heliocentric distance, mean proper motion, age and composition, not counting that they are located in a limited region of the sky [Tru30, Mer81, vLe09]. Of course now one can investigate the three-dimensional structure of the Hyades [Per98, Rei18], look for outlier Pleiads with abnormal proper motions [Ham93, Sar14], quantify an age spread in NGC 2264 [Par00, Ven14] or refine the age determination of IC 2391 with the lithium depletion boundary method [ByN99, Sod14], but after the second data release of *Gaia* [Gai18'] the major contributors to uncertainty of iPMO masses are theoretical models at very young ages [Bar02] and, in the case of high-extinction star-forming regions, the quality of spectroscopic data from which effective temperatures and, therefore, luminosities are derived.

Besides, as well as giant exoplanets, brown dwarfs and low-mass stars, iPMOs in young open clusters and star-forming regions are overluminous with respect to substellar objects of the same mass, but much older, in the field (as a reference, the Sun is 4.6 Ga old). In a sense, they are in a pseudo-Hayashi track that never reaches the zero-age main sequence. As shown in Fig. 2, iPMOs younger than about 10 Ma have bolometric luminosities equal or greater than those of the lowest mass stars older than the Hyades (650 Ma [Mar18]).

The younger, the brighter. Moreover, the younger, the easier recognizing a bona fide cluster member. Since iPMOs share characteristics with other cluster stars and brown dwarfs, there is a number of youth features that make them different from field interlopers. Spectroscopically, young iPMOs display weaker alkali neutral resonance lines (K I, Rb I, Cs I; but Li I deserves a separate discussion: Section 1), stronger titanium and vanadium oxide and weaker hydride absorption bands, and more peaked $H$-band pseudocontinuum than field objects of the same effective temperature [PR16, Lod18 and references therein]. Besides, their spectra can also show H$\alpha$, which may be strong and/or broad, and other emission lines typical in accreting T Tauri stars and brown dwarfs [Ber89, Cab06b]. The SEDs of several young iPMOs show mid- (or even near-)infrared excess, which is a signpost of a circumsubstellar disc, from which there can be accretion (Section 4). Juvenile iPMOS ($t$ = 30-200 Ma) do not have infrared flux discs, but may also have a flux redistribution between $J$ (1.2 mum) and $WISE$ $W2$ (4.6 μm) passbands, while keeping a constant bolometric luminosity [Fah13, Fil15, Fah16, ZO17]. Usually iPMOs in young clusters have L spectral types ($T_{eff}$ < 2200 K), but the least massive and/or in juvenile clusters (i.e. the Pleiades) have T spectral types ($T_{eff}$ < 1300 K). However, a T dwarf in the Hyades is still a brown dwarf (e.g. the two early T-dwarf Hyads discovered by [Bou08] have masses of about 0.05 M$_{sol}$). Non-deuterium-burning substellar objects in the Hyades might have a Y spectral type.

There have been claims of iPMO detections in a number of young open clusters with a wide range of heliocentric distances, ages and extinctions:

- Chamaeleon cloud complex [Oasa99', LM08, Luh08a, Muz15]
- Collinder 69 open [ByN07a, Bay12]
- IC 348 and NGC 1333 in Perseus [Sch09, Bur09, Sch12, AdO13, Luh16, Esp17]
- Lupus 3 [Muz14]
- Orion Nebula Cluster [LR00', Luc01', Luc06, Wei09, Hil13, Ing14, Sue14, Fan16]
- Pleiades [Cas07, Bih10, Cas10, ZO14a, ZO14b, ZO18]
- ρ Ophiuchi cloud complex [Hai10, Mar10, Gee11, AdO12, Muz12, AdO13, CC15]
- Serpens Core [Spe2]
- σ Orionis open cluster [ZO00', ZO02b', ZO02c', ByN01, Mar01, Cab07a', Bih09, Lod09, Béj11, PR12, PR15, ZO17']
- Taurus-Auriga [Luh09a, EL17]
- Upper Scorpius [Lod07, Lod08, Lod11, Lod13, PR16', Lod18']

Béjar & Martín [BM17] compiled 82 spectroscopically-confirmed iPMOs in open clusters and star-forming regions. Of them, 30 belonged to σ Orionis, 20 to the Orion Nebula Cluster, 21 to Upper Scorpius, seven to ρ Ophiuchi, and four to Chamaeleon, Lupus, Taurus and the Pleiades. There are a few differences between the compilation in [BM17] and Table 1 in this review, such as Cha J110913-773344, S Ori 69 and, especially, S Ori 70 now being considered as interlopers [Bur04b, Cab07', Luh08b, SJ08, ZO08, PR15']. Besides, since the compilation by [BM17] there have been a few new discoveries.

Below, I briefly review four works that present the very latest results in four representative regions: Taurus-Auriga, σ Orionis, Upper Scorpius and the Pleiades. Each of them have their own pros and cons for iPMO searches: Taurus-Auriga is very young and relatively nearby ($t$ ~ 1 Ma, $d$ ~ 140 pc) and has a substantial proper motion, but also has high, variable extinction of up to 20 mag and occupies an extended region in the sky [Eli78, KH95, And05]; Upper Scorpius and Taurus have very similar heliocentric distances, projected size and mean total proper motions, but the former has a much lower extinction, which is counterweighted by an older age of about 10 Ma [Wal94, Pre02, Pec12]; except for the thin nebulae around Merope, Alcyone, and Electra-Celaeno-Taygeta, the Pleiades cluster is virtually free of extinction and slightly closer than Taurus and Upper Scorpius, but is also much older, of about 120 Ma, which makes their iPMOs intrinsically fainter [Her47, Sod93, Mel14]; σ Orionis is located much further than the other three regions ($d$ = 386 pc [Sch16']), but has an age intermediate between Taurus and Upper Scorpius ($t$ = 3 Ma [ZO02a', She04]), is much more compact than them and, because of the intense radiation emitted by the the eponymous central Trapezium-like system (which erodes the famous Horsehead Nebula [SD15]), the extinction towards the cluster is very low [Lee68].

Most of the other listed star-forming regions (Chamaeleon, Perseus, Lupus, Orion Nebula Cluster, Ophiuchus, Serpens) have also high extinction and their iPMO candidates need a proper de-reddening. Only the Collinder 69 open cluster, in the Hunter Head, has a low extinction, probably due to the supernova explosion that originated the 12-deg ring around Meissa (λ Orionis) but, with $t$ = 5-10 Ma, it is slightly older than σ Orionis [Cow79, DM01, ByN04].

3.1.1. Taurus-Auriga

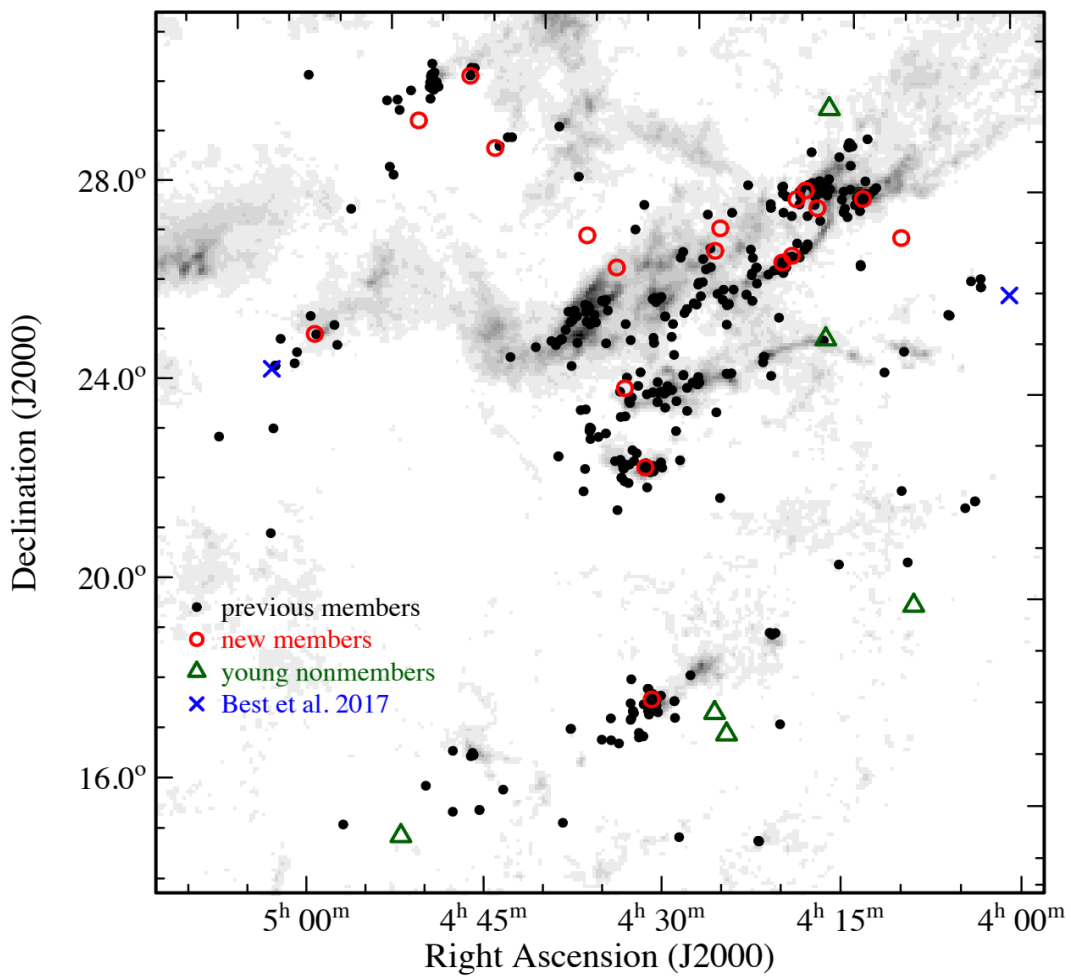

**Figure 6.** Spatial distribution of known members in Taurus. Black filled circles: members known before 2017; red open circles: new members from [EL17']; blue crosses: new members from [Bes17a]; green open triangles: young objects that do not appear to be members; grey scale: extinction map of [Dob05]. Reproduced with permission from [EL17'].

There are planets in Taurus: two of them have been imaged at wide physical separation of FW Tau AB and 2M0441+2301 B [BH15, Cac15] (see also FU Tau B, a 0.015 $M_{sol}$ brown dwarf at the deuterium-burning mass limit and companion of FU Tau A, a brown dwarf about three times more massive [Luh09b]). There are as well non-deuterium-burning substellar objects with similar masses, $M \sim 0.008\text{-}0.010$ $M_{sol}$, but free-floating in filaments in the intracluster medium together with stars and brown dwarfs (Fig. 6).

The latest and perhaps most comprehensive search for iPMOs in Taurus was done by Esplin & Luhman [EL17']. They used astro-photometric data from SDSS [York00], IRAC/*Spitzer* [Faz04], 2MASS [Skr06], UKIDSS [Law07], Pan-STARRS1 [Kai10], *WISE* [Wri10] and *Gaia* DR1 [Gai16]. First, they updated the list of Taurus members of [Luh17] with the new candidates proposed by [Kra17] and [Bes17a]. Next, they identified the best candidates with color-magnitude diagrams and proper motions, and obtained near-infrared spectroscopy of the most promising iPMOs candidates with SpeX/IRTF. After spectral classification (M9-L2), their sample included the four faintest known Taurus members and eight of the ten faintest ones in extinction-corrected $K_s$ band. Their least luminous targets should have masses as low as 0.004-0.005 $M_{sol}$ according to evolutionary models, and at least two of them have red mid-infrared colors relative of photospheres of young diskless objects of the same spectral type, which they ascribed to flux excesses from circum(sub)stellar disks (in the same work, [EL17'] also obtained spectra of low-mass brown dwarfs and iPMOS in IC 348 and NGC 1333 clusters in Perseus).

3.1.2. σ Orionis

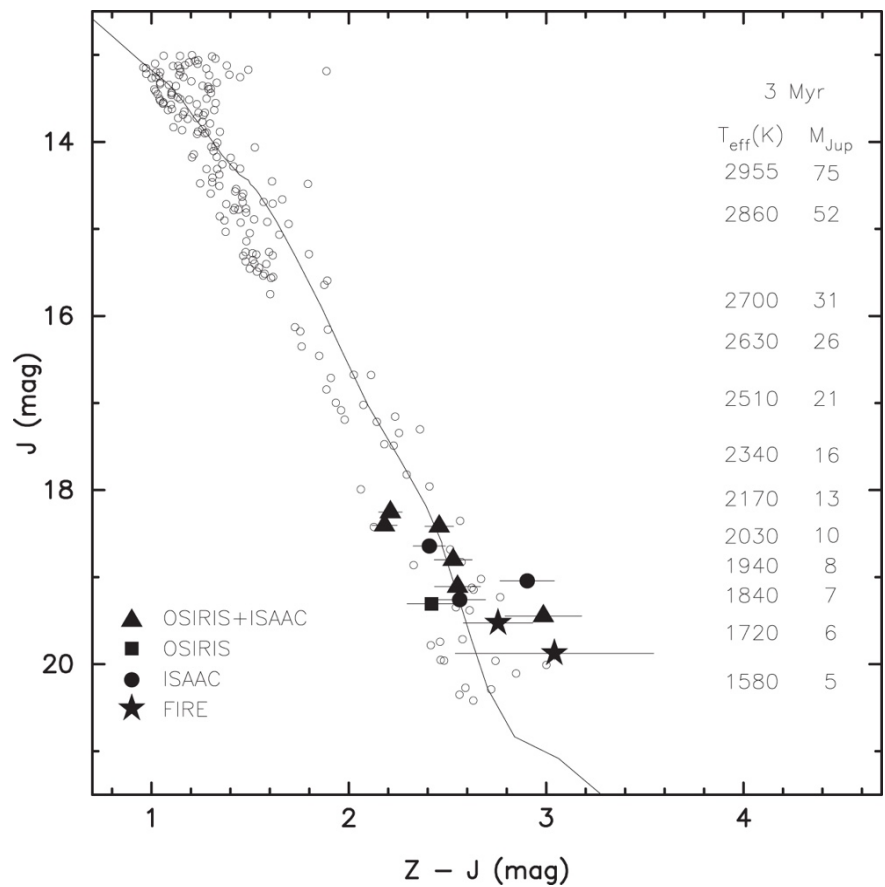

**Figure 7.** Optical/near-infrared color–magnitude diagram of σ Orionis low-mass star, brown dwarf, and iPMO candidates from [PR12'] (open circles) and [ZO17'] (filled symbols, with new spectroscopic follow-up; see legend). Solid line: 3 Ma isochrone [Cha00']. Columns: theoretical effective temperatures and masses (in $M_{Jup}$). Reproduced with permission from [ZO17'].

The σ Orionis open cluster in Ori OB1b is the the region *par excellence* for iPMO searches: it is very young, extinction-free, relatively close, and compact (the spatial distribution of bona fide members follow a stepped power-law radial profile up to 20 arcmin from the cluster centre [Cab08c] – in this core there is no second radial-velocity population as described by [Jef06], but there is overlapping with other younger stellar populations near the Horsehead Nebula and Alnitak/Flame Nebula at more than 30 arcmin [Cab08d]).

The work of Zapatero Osorio et al. [ZO17'] was the culmination of a two-decade effort in the search for the mass limit to formation via the opacity-limited fragmentation in σ Orionis. Already ten years earlier, [Cab07a'] had found that such limit must lie below 0.006 $M_{sol}$, but the at that time very few iPMOs had spectroscopy, and even less had clear signposts of youth. To fill this gap, [ZO17'] compiled the largest collection of high-quality spectra of the least massive objects in the cluster. In particular, they obtained low-resolution spectroscopy in the red optical and near infrared of 12 iPMOs with magnitudes *J* = 18.2-19.9 mag. With Osiris/GTC, ISAAC/VLT and FIRE/Magellan, they derived spectral types L0-4.5 and M9-L2.5 in the optical and near infrared, respectively, which correspond to effective temperatures of 2350-1800 K (for low surface gravities log g ~ 4.0). The targets spectra revealed signposts of youth (Section 3.1), thus "*corroborating their cluster membership and planetary masses*" (0.006-0.013 $M_{sol}$). Including six previously known σ Orionis L dwarfs in the spectro-photometric cluster sequence, two of which have disks [ByN01', Mar01', Luh08b'], "*these observations complete[d] the σ Orionis mass function by spectroscopically confirming the planetary-mass domain to a confidence level of ~75 %*". Their concluding remark, if true, will have an impact on future iPMO searches in the field (Section 6): [ZO17'] expected as many 0.006-0.0013 $M_{sol}$ iPMOs as 0.075-0.15 $M_{sol}$ late M and early L stars in the solar neighborhood, and they "*will evolve and look like the 2 pc-distant WISE J085510.83-071442.5* [Luh14a'] *at an age of a few gigayears*".

3.1.3. Upper Scorpius

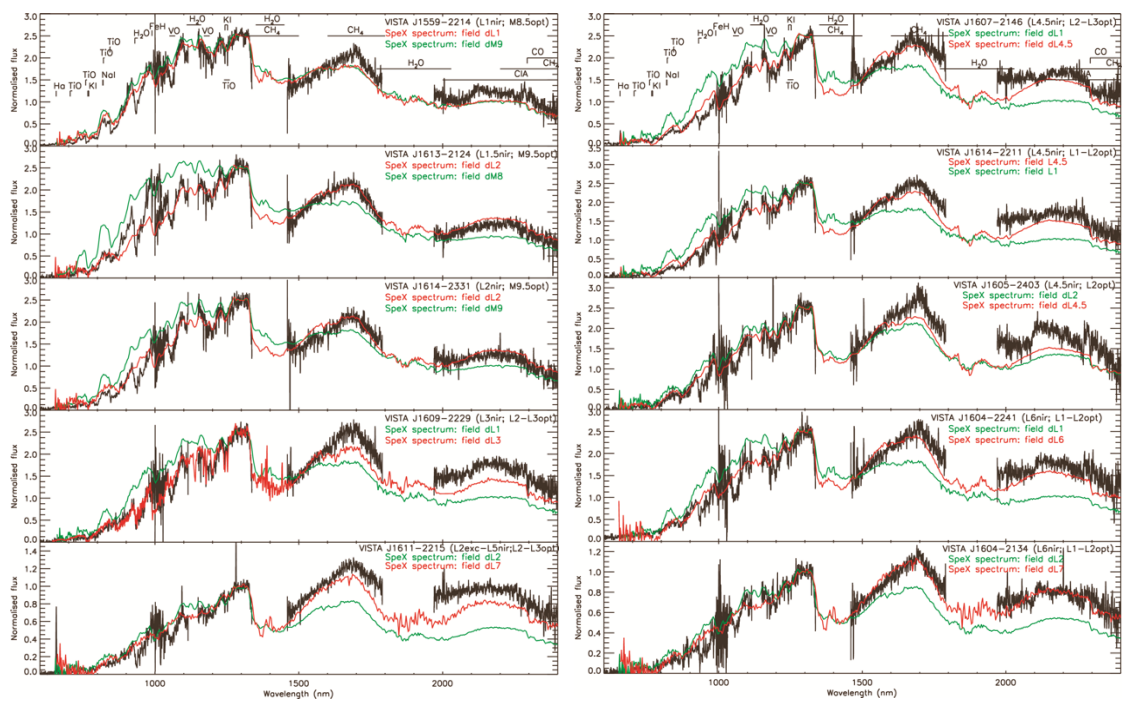

**Figure 8.** Normalized Osiris/GTC and *X*-shooter/VLT (smoothed) spectra of nine Upper Scorpius objects at the deuterium burning mass limit, compared to SpeX spectra of late M and early L field dwarfs. The overall spectral energy distribution matches best the infrared classification. Reproduced with permission from [Lod18'].

There are also planets in Upper Scorpius. Some of them are wide companions to stars and found in direct imaging, such as 1RXS J160929.1-210524 b [Laf08, Laf10] (GSC 06214-00210 B is likely a brown dwarf companion [Ire11, Lac15], such as HD 143567 B [Laf11]), while others are close companions to stars and found with *Kepler* and the transit method [Bor10, Lis14]. Of them, K2-33 b, a super-Neptune transiting an M3 Upper Scorpius low-mass star, is a cornerstone for understanding the formation and evolution of planets [Dav16, Man16, Van16].

Based on a previous photometric survey for objects in the planetary-mass domain, Lodieu et al. [Lod18'] presented the most exhaustive characterization of iPMOs in Upper Scorpius. The basis of their work was a very deep 13.5 deg² *ZYJ* VIRCAM/VISTA survey complemented with *ZYJHK* UKIDSS Galactic Cluster Survey data sets and *z*-band IMACS/Magellan imaging [Lod13']. In their new work, [Lod18'] used very deep *i*-band Osiris/GTC imaging and optical and near-infrared spectroscopy with Osiris/GTC and X-shooter/VLT, together with *WISE* mid-infrared photometry (and EMIR/GTC near-infrared spectroscopy of one target). Thanks to the corresponding analysis of the photometric and spectroscopic properties of young L-type Upper Scorpius members, they defined the first sequence of iPMOs in the association. Since their survey was limited by the *Y*-band depth, their *J*-band images might contain "some yet-to-be-found T-type dwarfs" with masses below 0.005 $M_{sol}$.

3.1.4. Pleiades

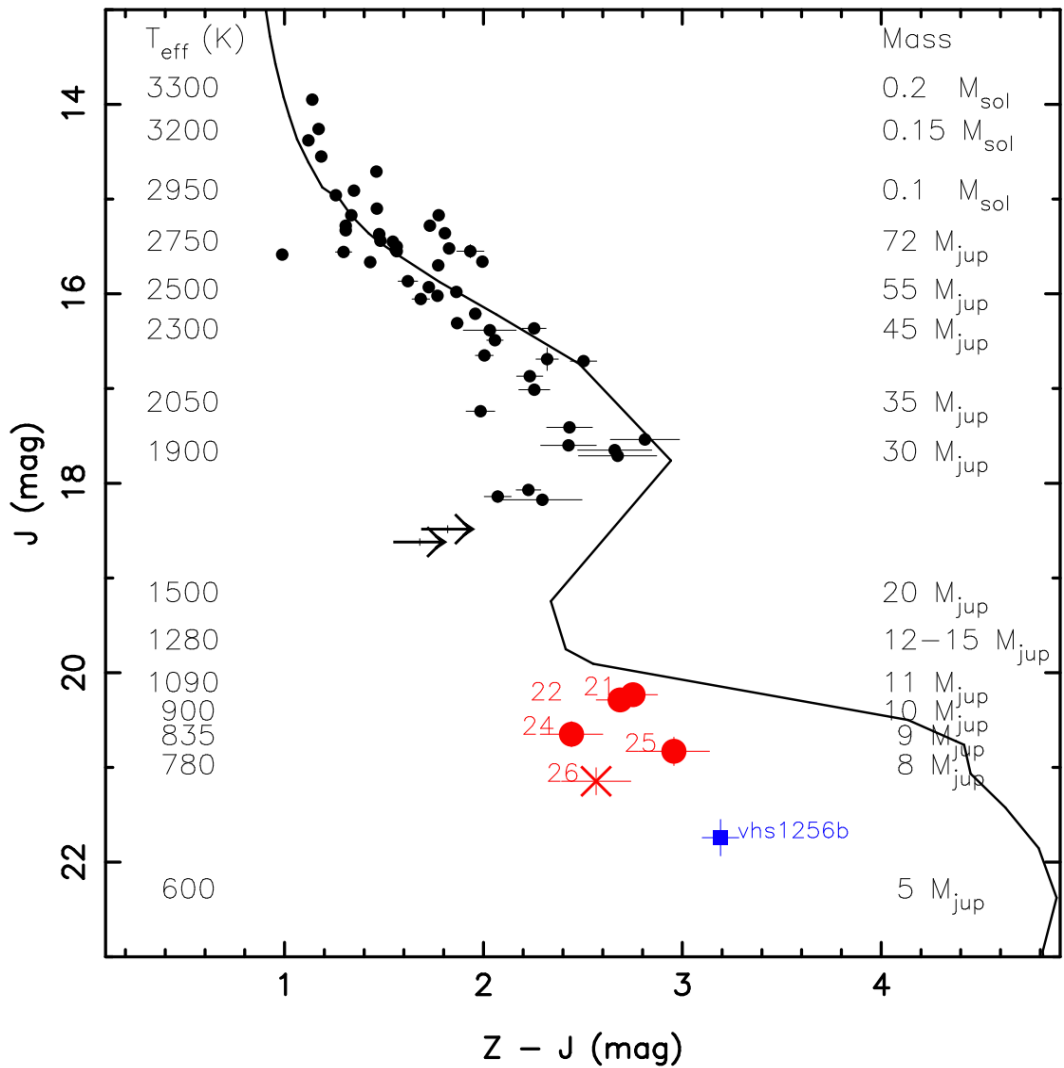

**Figure 9.** Optical/near-infrared color–magnitude diagram of Pleiades low-mass star, brown dwarf, and planetary-mass candidates from [Bih10', ZO14a'] (black dots) and [ZO18] (red filled circles; see legend). Red asterisk: likely non-member Calar 26. Black arrows: color from Z-band upper limits. Solid line: 120 Ma isochrone [Cha00']. Columns: theoretical effective temperatures and masses (in M$_{Jup}$). Blue square: VHS J1256-1257 b [Gau15] shifted to the Pleiades distance. Compare with Fig. 7. Reproduced with permission from [ZO18'].

The Pleiades (Messier 45, Seven Sisters, Siete Cabritillas, Siebengestirn, Sette Palommielle, daughters of Pleione, Manilius' narrow cloudy train of female stars, Bayer's Signatricia lumina et septistellium vestis insistoris, Hesperides, Al Thurayya/Al Najm, Subaru/Mutsuraboshi, Baba/Nasedha, Matariki, Krittika, Kimah, Cajupal, Sar en, Ulgher, Πλειαδες, [mao], $^{MUL}$MUL) was the first open cluster known to host substellar objects (Teide 1, PPl 15 AB [Reb95', Bas96']). However, it took almost a decade to find the first objects with masses below the deuterium-burning limit [Mar98, Bih06, Cas07', Lod12], including the ambiguous detection of T-dwarf candidates [Cas11]. In a survey that covered only about 3% of the total area of the cluster, [ZO14a'] discovered for the first time a population of 19 Pleiades iPMO candidates with proper motions consistent with cluster membership.

Using input data from [ZO14a'], in their new work, Zapatero Osorio et al. [ZO18'] imaged six Pleiades iPMO candidates in *z* band with Osiris/GTC: Calar 21, 22, 23, 24, 25 and 26. With the same instrument, they also imaged Calar 24 in *i* band, and took low-resolution (R ~ 270) optical spectroscopy of Calar 21 and 22. Because of its extended full width at half maximum with respect to point-like sources, Calar 23 is a background galaxy. Besides, the proper motion of Calar 26, the faintest source in *J* band, is compatible with that of the cluster at the 2.5 σ level, so only Calar 21, 22, 24 and 25 remain as astro-photometric Pleiad candidates. Of them, Calar 21 and 22, with approximate spectral types L6-7, extrapolate the spectro-photometric cluster sequence down to the deuterium-burning mass limit. The other two candidates, Calar 24 and 25, may have masses as low as 0.011-0.012 M$_{sol}$. The [ZO18'] findings demonstrated that not all iPMOs have completely escaped the Pleiades, in spite of their very low mass and cluster dynamical relaxation [Jam02, ZO14a']. Besides, the four Calar iPMOs are excellent targets for the deuterium test, which was proposed already by [Béj99'] to discriminate between brown dwarfs and "free-floating, non-deuterium-burning, substellar objects".

*3.2. (i)PMOs around brown dwarfs*

The 2003 WGESP definition of star, brown dwarf, planet and iPMO (Section 2) did not include the possibility of "non-deuterium-burning substellar objects" in orbit to brown dwarfs. A few such possible "planetary-mass companions to brown dwarfs" have been proposed (not the use of "B" instead of "b" in the name):

- TWA 27B (widely known as 2M1207-39b or 2M1207b) [Cha04', Cha05', Mam05, Son06, Duc08, Pat10, Bar11, Zho16]. It is the ~0.005 M$_{sol}$-mass common proper motion companion to the young brown dwarf TWA 27A in the ~10 Ma-old TW Hydrae association. They are separated by about 40 au and, because of the low mass of the primary, the system mass ratio is as low as 0.1-0.2, with a moderately high uncertainty that comes from the determination of the actual masses of both TWA 27 A and B. Furthermore, for explaining an apparent underluminosity of the secondary, it may have a surrounding disk, as well as the system primary, which has an impact on the derived mass [BC07, Moh07, MM07, Ske11, Ria12, Ric17].
- S Ori 68 [Cab06c]. Previously classified as an ~0.005 M$_{sol}$-mass iPMO in σ Orionis [ZO00', ByN01'], it lies at a projected physical separation of ~1700 au to the X-ray-flaring brown dwarf SE 70 [SE2004, Fra06]. In spite of being much more separated than TWA 27AB, [Cab06c] showed that the probability of chance alignment between the two cluster bodies was extremely low. Because of the system faintness ($J_A$ = 15.27 mag, $J_B$ = 20.2 mag), it misses a common proper motion confirmation. The corresponding system mass ratio is 0.2.
- L Ori 167 B [ByN07b]. It is a slightly older S Ori 68+SE 70 analog in the Collinder 69 ( λ Orionis) open cluster. Although L Ori 167 AB is poorly characterized, the two pairs share similar properties (location in Orion, projected physical separation of ~2000 au, secondary mass of ~0.008 M$_{sol}$). In the case of L Ori 167, the system mass ratio could be as high as 0.5.

- UScoCTIO 108 B [Béj08]. It is another wide ($s \sim 670$ au) companion to a young brown dwarf, in this case a high-mass one in Upper Scorpius. However, UScoCTIO 108 B could be as massive as 0.014 $M_{sol}$, which would disqualify it as a planetary-mass object. The system mass ratio is about 0.2.
- 2MASS J0441489+2301513 B (also known as 2M0441+23Bb) [Tod10, Tod14, BH15']. With a mass of 0.010 $M_{sol}$, it is the lowest-mass member of a young, hierarchical, quadruple system in Taurus containing a low-mass star, two brown dwarfs and the planetary-mass object. The primary 2M0441+23Ba at only 15 au is an M8.5V low-mass brown dwarf, and the pair is located at 1800 au to the more massive pair 2MASS J04414565+2301580 AB (2M0441+23AaAb). The Bb/Ba mass ratio is 0.5 [BH15'].

(2MASS J16222521-2405139 B, also known as Oph 11 B [JI06, Bra06, Clo07], is a low-mass brown dwarf companion to another higher mass brown dwarf, perhaps in Upper Scorpius [Luh07]). All five planetary-mass companion candidates above have early to mid L spectral types [ByN01', Cha05', BH15'], except for UScoCTIO 108 B, which is an M9.5V [Béj08']. Besides, L Ori 167 B has not been spectroscopically investigated yet, but its effective temperature estimated from photometry is 1750 K [ByN07b], well within the L domain. With ages younger than 10 Ma approximately, the five systems are also very young.

Even if they are gravitationally bound today, wide "brown dwarf+(i)PMO" systems in star-forming regions such as S Ori 68+SE 70, L Ori 167 AB or UScoCTIO 108 AB may not survive the interactions with other more massive cluster members and be disrupted in relatively short time scales (see [Cab09] and references therein). Furthermore, regardless TWA 27AB (2M1207-39b) and 2M0441+23Bb will eventually survive disruption by the Galactic gravitational potential, too, the mass ratio of the five systems, $M_B/M_A \sim 0.1$-$0.5$, is very high. Indeed, they look like a "petite version" of a brown dwarf (or stellar) binary formed through cloud fragmentation [Lod05] instead of a typical radial-velocity or transiting stellar system.

For complementing the 2003 definition, at the time of writing these lines WGESP has proposed several definition amendments to be voted during the 30th General Assembly of the International Astronomical Union (21-30 Aug 2018, Vienna, Austria). One of them is applying the term "planets" only to objects that have a mass ratio to the central object below the limiting ratio for stability of the triangular Lagrangian L4 and L5 points, i.e. $M_B/M_A < 2/(25+\sqrt{261}) \sim 0.02$. This mass ratio draws a boundary between two widely-separated groupings: stellar binaries and planets orbiting stars. With this new definition amendment, the objects listed above must not be considered planets, and become instead "brown dwarf-companion, non-deuterium-burning, substellar objects". Therefore, no planet around a brown dwarf has been detected yet (mid-L-type brown dwarfs have a shrinking habitable zone for those to-be-detected planets; T-type brown dwarfs are so faint that their theoretical habitable zone is inside the Roche radius [Des99, CR02, Cab10']).

*3.3. iPMOs in our vicinity*

Either they are companions to stars (or brown dwarfs) or are "really isolated", non-deuterium-burning substellar objects do exist in very young star-forming regions and open clusters. Gigayears later, when clusters are evaporated within the Galactic disk and iPMOs have cooled down to effective temperatures typical of Y dwarfs ($T_{eff} < 500$ K [Del08, Cus11, Kir11, Kir12, Liu12, Tin12, Bei13, Kir13, DK13, Cus14, Pin14]), they should be "just out there". Actually, TWA 27AB in TW Hydrae is located at merely about 55 pc. For that reason, it was not a great surprise when the first iPMOs in the solar neighborhood were reported as members in young kinematic associations [Liu13, Mac13]. More iPMOs arrived afterwards, especially in the 10 Ma-old TW Hydrae and 20 Ma-old beta Pictoris associations [Gag14a, Gag14b, Gag15a, Gag15b, Kel15, Sch16, Bes17b, Gag17a, Sch18b] and the slightly older AB Doradus and Carina-Near moving groups [Gag17b, Gag18]. See [Fah16'] for a compilation of young ultracool (spectral type > M7 V) member candidates in moving groups. Besides, CFBDSIR J2149-0403 was presented as an isolated planetary-mass object in AB Doradus [Del12], but to date it is not known whether it is a high-metallicity low-mass brown dwarf or a young iPMO not associated to any moving group [Del17].

Finally, *WISE*, with its astro-photometric capability in the mid infrared, also discovered a population of not-so-young iPMOs at very close heliocentric distances. These objects have so cool effective temperatures and the $CH_4$ absorption at 3.4 μm is so deep that they had escaped all earlier near-infrared all-sky surveys. They are so faint that they had not been able to characterize them spectroscopically in the near infrared until recently [Skemer16, ZO16]. The most extreme cases are perhaps L 97-3 B (WD 0806-661B) [Luh11, Rod11, Luh12, Luh14b] and, especially, WISE J085510.83-071442.5 [Luh14a', Kop14, LE14, Tin14, Wri14], which is the fourth closest stellar or substellar system and the coolest Y dwarf found to date (it has water ice clouds on the upper cloud layers; see [Yates17] for a discussion on atmospheric habitable zones in Y dwarfs). At $d$ = 2.41+0.08 pc [LE14] and the age of the Sun, WISE J085510.83-071442.5 would be an 0.005 $M_{sol}$ iPMO [ZO16].

To sum up, free-floating, substellar objects of a few Jupiter masses are indeed "just out there".

## 5. Formation

It was Shiv S. Kumar himself who proposed in 1963 for the first time, based on theoretical assumptions, that brown dwarfs (and objects beyond the deuterium burning mass limit) share the same formation mechanism [Ku64, Kum03']. He was also the first one to conclude that "*the mass of a gaseous fragment, formed by the star formation process, is much smaller than the minimum hydrogen burning mass*" [Kum67, Kum03']. This mass is what we call now opacity-mass limit, and has been settled at about 0.005 or slightly below $M_{sol}$ [LLB76, Ree76, Sil77, Toh80, Bos89, Bos01, Bat03].

[Whi07] reviewed from a theoretical perspective the five non-mutually-exclusive mechanisms for forming brown dwarfs, which are (1) turbulent fragmentation of molecular clouds, producing very-low-mass prestellar cores by shock compression; (2) collapse and fragmentation of more massive prestellar cores; (3) disk fragmentation; (4) premature ejection of protostellar embryos from their natal cores; and (5) photoerosion of pre-existing cores overrun by H II regions. [Luh07] complemented the [Whi07] review from a theoretical perspective. These mechanisms can in principle be extrapolated to iPMOs, together with the standard paradigm for the formation of giant planets in the Solar System plus subsequent ejection through very efficient dynamical interactions [Bos00, RC01, Bat02]. This standard planet-formation paradigm was reviewed by [MB98] and has six stages: (a) infall of dust grains onto the disk, evaporation and condensation; (b) formation and growth of solid particles from millimeter to kilometer sizes; (c) runaway coagulation of planetesimals into prototerrestrial planets; (d) concurrent accretion by gap formation in the disk; (e) termination of accretion by gap formation in the disk; and (f) clearing of the disk material.

By 2006, observational astronomers still looked for planet-like-formed iPMOs, which may have different internal structure (inner rocky core), composition (higher metallicity) and kinematics and position (ejected from birth system at high velocity). However, in that year Caballero concluded in his PhD thesis that "*very low-mass stars, brown dwarfs and iPMOs share the same formation mechanism [via turbulent fragmentation]*" (see [Cab07a', Cab10']). While the other mechanisms must not be ruled out completely and "*their relative importance probably depends on environment*" [Whit07], the standard mechanism of low-mass star formation can be extrapolated in general to brown dwarfs and iPMOs because of the following reasons:

- Continuity in the mass function. Even if it is described by a power law [Kro01'] or a log-normal function [Cha03'], the mass function (or the mass spectrum) in the low-mass stellar domain extrapolates smoothly to the substellar domain down to about 0.005-0.006 $M_{sol}$ [Cab07a', PR12', Muz15', EL17', Lod18'].
- Continuity in the frequency of discs. Isolated PMOs have discs, from which they accrete [ZO02b', Luh08b']. [ZO07'] measured for the first time the frequency of inner discs of objects between 0.007 and 0.014 $M_{sol}$. The observed rate in σ Orionis, greater or equal than 50%, was consistent with the rates measured for cluster brown dwarfs and low-mass stars, but suggested that "*there is a trend for the inner rate to increase with decreasing mass, which may be due to a mass-dependent timescale for the dissipation of the interior discs*" [ZO17'].

- Isolated PMOs, brown dwarfs and low-mass stars have the same spatial distribution [cf. Cab07a', PR12', Muz15', EL17', Lod18'] (but high-mass stars tend to be more concentrated towards the center of radially-symmetric clusters [Cab08c', Par14]). The lack of iPMOs in the very center of the σ Orionis cluster, near the Trapezium-like system, could be real or just an observational bias [Cab07b, Bou09].
- Proplyds, globulettes, proto-brown dwarfs and "Class 0 iPMOs" with substellar masses derived from radio and millimetric observations share the same properties as proto-stellar cores in extremely young star-forming regions [Pal12, Gah13, Pal14, Haw15, Mor15, dGM16, Liu16, Bay17, Ria17].

The star-like formation of iPMOs via turbulent fragmentation is now widely accepted as the most important, although not unique, scenario for explaining the formation of free-floating, non-deuterium-burning, substellar objects. However, regardless how they form, the fate of an iPMO is always the same: cool and cool for ever.

## 6. Future

Apart from cooling until Universe heat death, the mid- and long-term future of "free-floating, non-deuterium-burning, substellar objects" or iPMOs is summarized in three words: LSST, *Euclid*, *WFIRST*. The Large Synoptic Survey Telescope [LSS09], ESA *Euclid* mission [Lau11] and NASA *Wide Field Infrared Survey Telescope* [Spe15], with their large apertures and étendues in the optical and, especially, near infrared, will expand the sample of iPMOs with parallax determination in the solar vicinity, with proper-motion determination in open clusters and associations, and with accurate spectral energy distributions at all distances.

In the meantime, surveys such as BASS-Ultracool [Gag15b'], which aims at detecting T-type, planetary-mass members of young moving groups in the solar neighborhood, or SONYC [Muz15'], which provides a census of Substellar Objects in Nearby Young Clusters down to about 0.005 $M_{sol}$, will try to reach the opacity mass limit. Other key issues that will become very common in iPMO searches and analyses, and that need further development, will be narrow-band photometry [ZO18', Dea18], for which 8-10 m-class telescopes would work in survey mode, and improved theoretical models at very low temperatures and young ages [Bar02']. Beyond 2021, we will also use the *James Webb Space Telescope* [Bur03b, Tre17, Mor18] and the next generation of 30-40 m-class ground telescopes, such as the E-ELT [GS07], for the spectroscopic follow-up in the near and mid infrared of the faintest iPMOs.

A concluding science-fiction remark: because of their closeness, the coolest iPMOs may be eventually useful for mankind. What about mining their atmospheres in the far future? The atmosphere will be unbreathable, winds will be hurricane-force, surface gravity will be 100 times greater than on Earth (which implies difficulty for spaceflight [Hip18, How18]), but temperature will be comfortable and our descendants may be able to extract valuable ores demanded in the Solar System. Views from the iPMO terminator will be magnificent!

## 7. Postlude

There are examples of musical astronomy and astronomical music [Fra06, Ula09, Cab10, Lub10, Fra12]. As an unexpected end of review, I will show two examples of music and iPMOs. One is the videoclip of *El ordenador simula el nacimiento de las estrellas* (*Computer simulates birth of stars*), a song in Antonio Arias' album *Multiverso* that displays a reappraisal of the M. Bate's hydrodynamical modelling of the collapse and fragmentation of a cold, isothermal cloud, resulting in more than 1250 stars, brown dwarfs and iPMOs [mul, Bat12, Cab17].

The other example was "composed" by F. M. Walter in May 2002 in Hawai'i. Inspired by the discussion during the International Astronomical Union Symposium 211 *Brown Dwarfs* [Mar03] that resulted in the WGESP definition in Section 2 [Bos03], Fred prepared the lyrics below, to be accompanied of Woody Guthrie's *Talking Blues* (a talking blues is a strict-rhythm, free-melody, near-speech form of American folk/country music). Fred entitled the new song *The Brown Dwarf Talking*

*Blues* but, as you will read (or sing) below, he could instead have entitled it *The Free-Floating Non-Deuterium-Burning Substellar Object Talking Blues*.

> *Soaking up the rays at Waikoloa,*
> > *Two years before two thousand four,*
> *Pondering problems with nomenclature*
> > *Of heavenly orbs of tiny stature.*
> > > *Brown Dwarfs... Magenta Midgets...*
>
> *After dinner I attended a session,*
> > *hoping to learn a useful lesson.*
> *The big kahunas, and the Boss-man too*
> > *delivered the opinion of the IAU.*
> > > *Gas Giants... Sub-brown dwarfs...*
>
> *Things that fuse in the night are stars,*
> > *And orbiting them are planets like Mars.*
> *Can't see those cause of their low mass,*
> > *So we argue about great balls of gas.*
> > > *Free Floaters... Superplanets...*
>
> *Observations show they're free in space,*
> > *Theory says they must have come from some place.*
> *So what do you call that Jovian ball*
> > *Floating in space and not in thrall?*
> > > *Substellar mass objects... Plamos...*
>
> *Nature or nurture was the question to some,*
> > *Others just cared for the mass, by gum!*
> *Political correctness carried the day:*
> > *Tally up the names in the papers, they say.*
> > > *Mass-challenged stars... russet runts...*
>
> *If you ask me it doesn't make much sense*
> > *To hotly debate our ignorance.*
> *Seems to me planets are really obscene...*
> > *When you see it you'll know it, if you get what I mean.*
> > > *Damn Degenerates...*

I echo the lyrics of Fred's talking blues with my last digression. When the time comes with improved knowledge on iPMOs, the hot debate on their nomenclature will be back. As a preparation for that moment, I present a new old point of view: Latin. In the classical "language of international communication, scholarship and science until well in the 18th century", brown dwarf is *pumilio fusca*. However, brown dwarfs are not actual dwarfs because they never reach the main sequence; this fact takes on relevance especially at very young ages, when coeval stars above the hydrogen burning mass limit and very low surface gravities belong to spectroscopic classes different from V. Since we should avoid the terms *stella*, *astrum* and *aster* (astro-, star) for brown dwarf, we could use instead other Latin term for designating heavenly bodies in general: *sidus* (sideral). In Table 2, I compiled the color terms for "brown" in Botanical Latin; *fuscus* is greyish brown. If we use instead *rufus*, red brown and keep the root of *sidus* without its proper declension (third, neuter), we get a funny word for brown dwarf: *siderufo*. If we use the prefix *hypo-* from Ancient Greek instead of *sub-* from Latin for indicating that they are "under brown dwarfs", and put everything together, my naming proposal for "free-floating, non-deuterium-burning, substellar object" (iPMO) gets *hyposiderufo*.

**Table 2.** Color terms for "brown" in Botanical Latin [bot].

| Botanical Latin | Meaning |
| --- | --- |
| *badius* | chestnut brown |
| *boeticus* | Spanish brown |
| *brunneus* | pure dull brown |
| *cacainus* | chocolate brown |
| *chocolatinus* | chocolate brown |
| *cinnamomeus* | cinnamon |
| *coffeatus* | coffee-bean brown |
| *cupreus* | brownish red |
| *ferrugineus* | rusty brown |
| *fuligineus* | sooty brown |
| *fuliginosus* | sooty brown |
| *fuscus* | greyish brown |
| *glandaceous* | yellowish red brown |
| *haematiticus* | brown red |
| *hepaticus* | liver brown |
| *ligneus* | wood brown |
| *luridus* | cloudy brown |
| *nicotanus* | tobacco leaf brown |
| *phaeo-* | greyish brown |
| *porphyreus* | reddish brown |
| *rubiginosus* | brown red |
| *rufescens* | red brown |
| *rufus* | red brown |
| *sanguineus* | dull red, brownish black |
| *spadiceus* | bright brown |
| *theobromius* | chocolate brown |
| *umbrinus* | umber brown |
| *ustalus* | charred wood brown |
| *vaccinus* | cow brown |
| *xerampelinus* | dull red with brown |

**Funding:** This research was funded by the Spanish Ministerio de Ciencia, Innovación y Universidades through project AYA2016-79425-C3-2-P.

**Acknowledgments:** I thank V. J. S. Béjar for providing me with useful material for this review and his patience for so many years, G. Chabrier, T. Esplin, N. Lodieu and M. R. Zapatero Osorio for kindly giving permission to reproduce figures, and M. Oshagh for offering me the chance to review this topic.

**Conflicts of Interest:** The author declares no conflict of interest.

**Note to the editor and referee(s):** I use SI symbol "a" (annus) for year. During refereeing, I keep the [MmmYY] acronym (w/o apostrophe) for the references; once accepted, I will use only numerals [N].

**Note to the arXiv reader:** This manuscript is being refereed: prompt comments and suggestions to caballero@cab.inta-csic.es are more than welcome.

**References**

1. [vB44] van Biesbroeck, G., The star of the lowest known luminosity, *AJ* **1944**, *51*, 61, DOI 10.1086/105801.
2. [Her56] Herbig, G. H., Observations of the Spectrum of the Companion to BD + 4°4048, *PASP* **1956**, *68*, 53, DOI 10.1086/126992.

3.  [Kir91] Kirkpatrick, J. D., Henry, T. J., McCarthy, D. W., Jr., A standard stellar spectral sequence in the red/near-infrared - Classes K5 to M9, *ApJS* **1991**, *77*, 417, DOI 10.1086/191611.
4.  [Gai18] Gaia Collaboration, Brown, A. G. A., Vallenari, A., Prusti, T., de Bruijne, J. H. J., Babusiaux, C., Bailer-Jones, C. A. L., Gaia Data Release 2. Summary of the contents and survey properties, *A&A* **2018**, *616*, A1, DOI 10.1051/0004-6361/201833955.
5.  [AF15] Alonso-Floriano, F. J., Morales, J. C., Caballero, J. A. et al., CARMENES input catalogue of M dwarfs. I. Low-resolution spectroscopy with CAFOS, *A&A* **2015**, *577*, A128, DOI 10.1051/0004-6361/201525803.
6.  [Kam18] Kaminski, A., Trifonov, T., Caballero, J. A. et al., The CARMENES search for exoplanets around M dwarfs. A Neptune-mass planet traversing the habitable zone around HD 180617, *A&A* **2018**, in press, eprint arXiv:1808.01183.
7.  [Sch18a] Schweitzer, A., Passegger, V. M., Béjar, V. J. S. et al., The CARMENES search for exoplanets around M dwarfs. The different roads to radii and masses of the target stars, *A&A*, in prep.
8.  [PL83] Probst, R. G., Liebert, J., LHS 2924 - A uniquely cool low-luminosity star with a peculiar energy distribution, *ApJ* **1983**, *274*, 245, DOI 10.1086/161442.
9.  [Kum03] Kumar, S. S., The Bottom of the Main Sequence and Beyond: Speculations, Calculations, Observations, and Discoveries (1958-2002). Brown Dwarfs, Proceedings of IAU Symposium #211, held 20-24 May 2002 at University of Hawaii, Honolulu, Hawaii. Edited by Eduardo Martín. San Francisco: Astronomical Society of the Pacific, **2003**, p. 3.
10. [Kum63] Kumar, S. S., The Structure of Stars of Very Low Mass, *ApJ* **1963**, *137*, 1121, DOI 10.1086/147589.
11. [HN63] Hayashi, C., Nakano, T., Evolution of Stars of Small Masses in the Pre-Main-Sequence Stages, *Progress of Theoretical Physics* **1963**, *30*, 460, DOI 10.1143/PTP.30.460.
12. [MR67] Mestel, L., Ruderman, M. A., The energy content of a white dwarf and its rate of cooling, *MNRAS* **1967**, *136*, 27, DOI 10.1093/mnras/136.1.27.
13. [Vil71] Vila, S. C., Evolution of a 0.6 $M_{sol}$ White Dwarf, *ApJ* **1971**, *170*, 153, DOI 10.1086/151196.
14. [Tar14] Tarter, J., Brown Is Not a Color: Introduction of the Term 'Brown Dwarf', 50 Years of Brown Dwarfs, Astrophysics and Space Science Library, Volume 401. ISBN 978-3-319-01161-5. Springer International Publishing Switzerland, **2014**, p. 19, DOI 10.1007/978-3-319-01162-2_3.
15. [Irw91] Irwin, M., McMahon, R. G., Reid, N., A star of exceedingly low luminosity, *MNRAS* **1991**, *252*, P61, DOI 10.1093/mnras/252.1.61P.
16. [Sch91] Schneider, D. P., Greenstein, J. L., Schmidt, M., Gunn, J. E., Spectroscopy of an unusual emission line M star, *AJ* **1991**, *102*, 1180, DOI 10.1086/115945.
17. [BZ88] Becklin, E. E., Zuckerman, B., A low-temperature companion to a white dwarf star, *Nature* **1988**, *336*, 656, DOI 10.1038/336656a0.
18. [Kir99a] Kirkpatrick, J. D., Allard, F., Bida, T. et al., An Improved Optical Spectrum and New Model FITS of the Likely Brown Dwarf GD 165B, *ApJ* **1999**, *519*, 834, DOI 10.1086/307380.
19. [Lat89] Latham, D. W., Mazeh, T., Stefanik, R. P., Mayor, M. Burki, G., The unseen companion of HD114762 - A probable brown dwarf, *Nature* **1989**, *339*, 38, DOI 10.1038/339038a0.
20. [McC84] McCarthy, D. W., Jr., Probst, R. G., Detection of an Infrared Source Near VB 8: The First Extra-solar Planet?, *Bulletin of the American Astronomical Society* **1984**, *96*, 165.
21. [ZB87] Zuckerman, B., Becklin, E. E., Excess infrared radiation from a white dwarf - an orbiting brown dwarf?, *Nature* **1987**, *330*, 138, DOI 10.1038/330138a0.
22. [LK75] Luyten, W. J., Kowal, C. T., Proper motion survey with the forty-eight inch Schmidt telescope. XLIII. One hundred and six faint stars with large proper motions, Separate print Univ. Minnesota, Minneapolis, Minnesota, **1975**, 2 p.
23. [Tin98] Tinney, C. G., The intermediate-age brown dwarf LP944-20, *MNRAS* **1989**, *296*, L42, DOI 10.1046/j.1365-8711.1998.01642.x.
24. [RR90] Rieke, G. H., Rieke, M. J., Possible substellar objects in the Rho Ophiuchi cloud, *ApJ* **1990**, *362*, L21, DOI 10.1086/185838.
25. [Com98] Comerón, F., Rieke, G. H., Claes, P., Torra, J., Laureijs, R. J., ISO observations of candidate young brown dwarfs, *A&A* **1998**, *335*, 522.
26. [Bas14] Basri, G., The Discovery of the First Lithium Brown Dwarf: PPl 15, 50 Years of Brown Dwarfs, Astrophysics and Space Science Library, Volume 401. ISBN 978-3-319-01161-5. Springer International Publishing Switzerland, **2014**, p. 51, DOI 10.1007/978-3-319-01162-2_5.

27. [MQ95] Mayor, M., Queloz, D., A Jupiter-mass companion to a solar-type star, *Nature* **1995**, *378*, 355, DOI 10.1038/378355a0.
28. [Reb95] Rebolo, R., Zapatero Osorio, M. R., Martín, E. L., Discovery of a brown dwarf in the Pleiades star cluster, Nature 1995, 377, 129, DOI 10.1038/377129a0.
29. [Reb96] Rebolo, R., Martín, E. L., Basri, G., Marcy, G. W., Zapatero Osorio, M. R., Brown Dwarfs in the Pleiades Cluster Confirmed by the Lithium Test, ApJ 1996, 469, L53, DOI 10.1086/310263.
30. [Bas96] Basri, G., Marcy, Geoffrey W., Graham, J. R., Lithium in Brown Dwarf Candidates: The Mass and Age of the Faintest Pleiades Stars, ApJ 1996, 458, 600, DOI 10.1086/176842.
31. [Nak95] Nakajima, T., Oppenheimer, B. R., Kulkarni, S. R. et al., Discovery of a cool brown dwarf, Nature 1995, 378, 463, DOI 10.1038/378463a0.
32. [Opp95] Oppenheimer, B. R., Kulkarni, S. R., Matthews, K., Nakajima, T., Infrared Spectrum of the Cool Brown Dwarf Gl 229B, Science 1995, 270, 1478.
33. [Béj99] Béjar, V. J. S., Zapatero Osorio, M. R., Rebolo, R., A Search for Very Low Mass Stars and Brown Dwarfs in the Young σ Orionis Cluster, ApJ 1999, 521, 671, DOI 10.1086/307583.
34. [ZO02a] Zapatero Osorio, M. R., Béjar, V. J. S., Pavlenko, Ya. et al., Lithium and Hα in stars and brown dwarfs of sigma Orionis, A&A 2002a, 384, 937, DOI 10.1051/0004-6361:20020046.
35. [Cab07a] Caballero, J. A., Béjar, V. J. S., Rebolo, R. et al., The substellar mass function in σ Orionis. II. Optical, near-infrared and IRAC/Spitzer photometry of young cluster brown dwarfs and planetary-mass objects, A&A 2007, 470, 903, DOI 10.1051/0004-6361:20066993.
36. [Luh13] Luhman, K. L., Discovery of a Binary Brown Dwarf at 2 pc from the Sun, ApJ 2013, 767, L1, DOI 10.1088/2041-8205/767/1/L1.
37. [Sma17] Smart, R. L., Marocco, F., Caballero, J. A. et al., The Gaia ultracool dwarf sample - I. Known L and T dwarfs and the first Gaia data release, MNRAS 2017, 469, 401, DOI 10.1093/mnras/stx800.
38. [Bur97] Burrows, A., Marley, M., Hubbard, W. B. et al., A Nongray Theory of Extrasolar Giant Planets and Brown Dwarfs, ApJ 1997, 491, 856, DOI 10.1086/305002.
39. [Bar98] Baraffe, I., Chabrier, G., Allard, F., Hauschildt, P. H., Evolutionary models for solar metallicity low-mass stars: mass-magnitude relationships and color-magnitude diagrams, A&A 1998, 337, 403.
40. [Cha00] Chabrier, G., Baraffe, I., Allard, F., Hauschildt, P., Evolutionary Models for Very Low-Mass Stars and Brown Dwarfs with Dusty Atmospheres, ApJ 2000, 542, 464, DOI 10.1086/309513.
41. [All01] Allard, F., Hauschildt, P. H., Alexander, D. R., Tamanai, A., Schweitzer, A., The Limiting Effects of Dust in Brown Dwarf Model Atmospheres, ApJ 2001, 556, 357, DOI 10.1086/321547.
42. [Bat03] Bate, M. R., Bonnell, I. A., Bromm, V., The formation of a star cluster: predicting the properties of stars and brown dwarfs, MNRAS 2003, 339, 577, DOI 10.1046/j.1365-8711.2003.06210.x.
43. [Bar15] Baraffe, I., Homeier, D., Allard, F., Chabrier, G., New evolutionary models for pre-main sequence and main sequence low-mass stars down to the hydrogen-burning limit, A&A 2015, 577, A42, DOI 10.1051/0004-6361/201425481.
44. [MR15] Marley, M. S., Robinson, T. D., On the Cool Side: Modeling the Atmospheres of Brown Dwarfs and Giant Planets, ARA&A 2015, 53, 279, DOI 10.1146/annurev-astro-082214-122522.
45. [Del99] Delfosse, X., Tinney, C. G., Forveille, T. et al., Searching for very low-mass stars and brown dwarfs with DENIS, A&AS 1999, 135, 41, DOI 10.1051/aas:1999158.
46. [Bur00] Burgasser, A. J., Kirkpatrick, J. D., Cutri, R. M. et al., Discovery of a Brown Dwarf Companion to Gliese 570ABC: A 2MASS T Dwarf Significantly Cooler than Gliese 229B, ApJ 2000, 531, L57, DOI 10.1086/312522.
47. [Giz00] Gizis, J. E., Monet, D. G., Reid, I. N. et al., New Neighbors from 2MASS: Activity and Kinematics at the Bottom of the Main Sequence, AJ 2000, 120, 1085, DOI 10.1086/301456.
48. [BJM01] Bailer-Jones, C. A. L., Mundt, R., Variability in ultra cool dwarfs: Evidence for the evolution of surface features, A&A 2001, 367, 218, DOI 10.1051/0004-6361:20000416.
49. [Bou03] Bouy, H. Brandner, W. Martín, E. L. et al., Multiplicity of Nearby Free-Floating Ultracool Dwarfs: A Hubble Space Telescope WFPC2 Search for Companions, AJ 2003, 126, 1526, DOI 10.1086/377343.
50. [WB03] White, R. J., Basri, G., Very Low Mass Stars and Brown Dwarfs in Taurus-Auriga, ApJ 2003, 582, 1109, DOI 10.1086/344673.
51. [ZO05] Zapatero Osorio, M. R., Caballero, J. A., Béjar, V. J. S., Optical Linear Polarization of Late M and L Type Dwarfs, ApJ 2005, 621, 445, DOI 10.1086/427433.

52. [Mor08] Morin, J., Donati, J.-F., Petit, P. et al., Large-scale magnetic topologies of mid M dwarfs, MNRAS 2008, 390, 567, DOI 10.1111/j.1365-2966.2008.13809.x.
53. [Art09] Artigau, É., Bouchard, S., Doyon, R., Lafrenière, D., Photometric Variability of the T2.5 Brown Dwarf SIMP J013656.5+093347: Evidence for Evolving Weather Patterns, ApJ 2009, 701, 1534, DOI 10.1088/0004-637X/701/2/1534.
54. [Rad12] Radigan, J., Jayawardhana, R., Lafrenière, D. et al., Large-amplitude Variations of an L/T Transition Brown Dwarf: Multi-wavelength Observations of Patchy, High-contrast Cloud Features, ApJ 2012, 750, 105, DOI 10.1088/0004-637X/750/2/105.
55. [DK13] Duchêne, G., Kraus, A., Stellar Multiplicity, ARA&A 2013, 51, 269, DOI 10.1146/annurev-astro-081710-102602.
56. [Kir99b] Kirkpatrick, J. D., Reid, I. N., Liebert, J. et al., Dwarfs Cooler than "M": The Definition of Spectral Type "L" Using Discoveries from the 2 Micron All-Sky Survey (2MASS), ApJ 1999, 519, 802, DOI 10.1086/307414.
57. [Bur02] Burgasser, A. J., Kirkpatrick, J. D., Brown, M. E. et al. The Spectra of T Dwarfs. I. Near-Infrared Data and Spectral Classification, ApJ 2002, 564, 421, DOI 10.1086/324033.
58. [Geb02] Geballe, T. R., Knapp, G. R., Leggett, S. K. et al., Toward Spectral Classification of L and T Dwarfs: Infrared and Optical Spectroscopy and Analysis, ApJ 2002, 564, 466, DOI 10.1086/324078.
59. [Kir05] Kirkpatrick, J. D., New Spectral Types L and T, ARA&A 2005, 43, 195, DOI 10.1146/annurev.astro.42.053102.134017.
60. [Kir12] Kirkpatrick, J. D., Gelino, C. R., Cushing, M. C. et al., Further Defining Spectral Type "Y" and Exploring the Low-mass End of the Field Brown Dwarf Mass Function, ApJ 2012, 753, 156, DOI 10.1088/0004-637X/753/2/156.
61. [Luh14a] Luhman, K. L., Discovery of a ~250 K Brown Dwarf at 2 pc from the Sun, ApJ 2014, 786, L18, DOI 10.1088/2041-8205/786/2/L18.
62. [Reb92] Rebolo, R., Martín, E. L., Magazzù, A, Spectroscopy of a brown dwarf candidate in the Alpha Persei open cluster, ApJ 1992, 389, L83, DOI 10.1086/186354.
63. [Mag93] Magazzù, A., Martín, E. L., Rebolo, R., A spectroscopic test for substellar objects, ApJ 1993, 404, L17, DOI 10.1086/186733.
64. [Sau96] Saumon, D., Hubbard, W. B., Burrows, A., A Theory of Extrasolar Giant Planets, ApJ 1996, 460, 993, DOI 10.1086/177027.
65. [Spi11] Spiegel, D. S., Burrows, A., Milsom, J. A., The Deuterium-burning Mass Limit for Brown Dwarfs and Giant Planets, ApJ 2011, 727, 57, DOI 10.1088/0004-637X/727/1/57.
66. [CB00] Chabrier, G., Baraffe, I., Theory of Low-Mass Stars and Substellar Objects, ARA&A 2000, 38, 337, DOI 10.1146/annurev.astro.38.1.337.
67. [Cha00] Charbonneau, D., Brown, T. M., Latham, D. W., Mayor, M., Detection of Planetary Transits Across a Sun-like Star, ApJ 2000, 529, L45, DOI 10.1086/312457.
68. [Cha02] Charbonneau, D., Brown, T. M., Noyes, R. W., Gilliland, R. L., Detection of an Extrasolar Planet Atmosphere, ApJ 2002, 568, 377, DOI 10.1086/338770.
69. [But04] Butler, R. P., Vogt, S. S., Marcy, G. W. et al., A Neptune-Mass Planet Orbiting the Nearby M Dwarf GJ 436, ApJ 2004, 617, 580, DOI 10.1086/425173.
70. [Bea06] Beaulieu, J.-P., Bennett, D. P., Fouqué, P. et al., Discovery of a cool planet of 5.5 Earth masses through gravitational microlensing, Nature 2006, 439, 437, 10.1038/nature04441.
71. [Knu07] Knutson, H. A., Charbonneau, D., Allen, L. E. et al., A map of the day-night contrast of the extrasolar planet HD 189733b, Nature 2007, 447, 183, DOI 10.1038/nature05782.
72. [Udr07] Udry, S.; Bonfils, X.; Delfosse, X. et al., The HARPS search for southern extra-solar planets. XI. Super-Earths (5 and 8 $M_{Terra}$) in a 3-planet system, A&A 2007, 469, L43, DOI 10.1051/0004-6361:20077612.
73. [Bat11] Batalha, N. M., Borucki, W. J., Bryson, S. T. et al., Kepler's First Rocky Planet: Kepler-10b, ApJ 2011, 729, 27, DOI 10.1088/0004-637X/729/1/27.
74. [Doy11] Doyle, L. R., Carter, J. A., Fabrycky, D. C. et al., Kepler-16: A Transiting Circumbinary Planet, Science 2011, 333, 1602, DOI 10.1126/science.1210923.
75. [Qui14] Quintana, E. V., Barclay, T., Raymond, S. N. et al., An Earth-Sized Planet in the Habitable Zone of a Cool Star, Science 2014, 344, 277, DOI 10.1126/science.1249403.
76. [Mac15] Macintosh, B., Graham, J. R., Barman, T. et al., Discovery and spectroscopy of the young jovian planet 51 Eri b with the Gemini Planet Imager, Science 2015, 350, 64, DOI 10.1126/science.aac5891.

77. [AE16] Anglada-Escudé, G., Amado, P. J., Barnes, J. et al., A terrestrial planet candidate in a temperate orbit around Proxima Centauri, Nature 2016, 536, 437, DOI 10.1038/nature19106.
78. [Gil17] Gillon, M., Triaud, A. H. M. J., Demory, B.-O. et al., Seven temperate terrestrial planets around the nearby ultracool dwarf star TRAPPIST-1, Nature 2017, 542, 456, DOI 10.1038/nature21360.
79. [MB96] Marcy, G. W., Butler, R. P., A Planetary Companion to 70 Virginis, ApJ 1996, 464, L147, DOI 10.1086/310096.
80. [But97] Butler, R. P., Marcy, G. W., Williams, E., Hauser, H., Shirts, P., Three New "51 Pegasi-Type" Planets, ApJ 1997, 474, L115, DOI 10.1086/310444.
81. [Mar08] Marois, C., Macintosh, B., Barman, T. et al., Direct Imaging of Multiple Planets Orbiting the Star HR 8799, Science 2008, 322, 1348, DOI 10.1126/science.1166585.
82. [Mar10] Marois, C., Zuckerman, B., Konopacky, Q. M., Macintosh, B., Barman, T., Images of a fourth planet orbiting HR 8799, Nature 2010, 468, 1080, DOI 10.1038/nature09684.
83. [Tsa18] Tsapras, Y., Preliminary topic: Microlensing searches for exoplanets, Geosciences, THIS VOLUME.
84. [bur] Professor Adam Burrows' Home Page, Astrophysics, Supernovae, Planets, Exoplanets. Available online: https://www.astro.princeton.edu/~burrows/ (accessed on 22 Aug 2018).
85. [Bos07] Boss, A. P., Butler, R. P., Hubbard, W. B., Working Group on Extrasolar Planets, IAU Transactions, Vol. 26A, Reports on Astronomy 2002-2005. Edited by O. Engvold. Cambridge: Cambridge University Press, 2007, p.183, DOI 10.1017/S1743921306004509.
86. [Bur03a] Burgasser, A. J., Kirkpatrick, J. D., Burrows, A. et al., The First Substellar Subdwarf? Discovery of a Metal-poor L Dwarf with Halo Kinematics, ApJ 2003, 592, 1186, DOI 10.1086/375813.
87. [Bur04a] Burgasser, A. J., Discovery of a Second L Subdwarf in the Two Micron All Sky Survey, ApJ 2004, 614, L73, DOI 10.1086/425418.
88. [Sch04] Scholz, R.-D., Lodieu, N., McCaughrean, M. J., SSSPM J1444-2019: An extremely high proper motion, ultracool subdwarf, A&A 2004, 428, L25, DOI 10.1051/0004-6361:200400098.
89. [Lod17] Lodieu, N., Espinoza Contreras, M., Zapatero Osorio, M. R. et al., New ultracool subdwarfs identified in large-scale surveys using Virtual Observatory tools, A&A 2017, 598, A92, DOI 10.1051/0004-6361/201629410.
90. [Zha17a] Zhang, Z. H., Pinfield, D. J., Gálvez-Ortiz, M. C. et al., Primeval very low-mass stars and brown dwarfs - I. Six new L subdwarfs, classification and atmospheric properties, MNRAS 2017, 464, 3040, DOI 10.1093/mnras/stw2438.
91. [Zha17b] Zhang, Z. H., Homeier, D., Pinfield, D. J. et al., Primeval very low-mass stars and brown dwarfs - II. The most metal-poor substellar object, MNRAS 2017, 468, 261, DOI 10.1093/mnras/stx350.
92. [exo] The Extrasolar Planets Encyclopaedia. Available online: http://exoplanets.eu (accessed on 22 Aug 2018).
93. [HR15] Hatzes, A. P., Rauer, H., A Definition for Giant Planets Based on the Mass-Density Relationship, ApJ 2015, 810, L25, DOI 10.1088/2041-8205/810/2/L25.
94. [Sal55] Salpeter, E. E., The Luminosity Function and Stellar Evolution, ApJ 1955, 121, 161, DOI 10.1086/145971.
95. [Sca86] Scalo, J. M., The stellar initial mass function, Fundamentals of Cosmic Physics, 11, 1, ISSN 0094-5846.
96. [Kro01] Kroupa, P., On the variation of the initial mass function, MNRAS 2001, 322, 231, DOI 10.1046/j.1365-8711.2001.04022.x.
97. [Cha03] Chabrier, G., Galactic Stellar and Substellar Initial Mass Function, PASP 2003, 115, 763, 10.1086/376392.
98. [LR00] Lucas, P. W., Roche, P. F., A population of very young brown dwarfs and free-floating planets in Orion, MNRAS 2000, 314, 858, DOI 10.1046/j.1365-8711.2000.03515.x.
99. [ZO00] Zapatero Osorio, M. R., Béjar, V. J. S., Martín, E. L. et al., Discovery of Young, Isolated Planetary Mass Objects in the σ Orionis Star Cluster, Science 2000, 290, 103, DOI 10.1126/science.290.5489.103.
100. [Luc01] Lucas, P. W., Roche, P. F., Allard, F., Hauschildt, P. H., Infrared spectroscopy of substellar objects in Orion, MNRAS 2001, 326, 695, DOI 10.1046/j.1365-8711.2001.04666.x.
101. [Oas99] Oasa, Y., Tamura, M., Sugitani, K., A Deep Near-Infrared Survey of the Chamaeleon I Dark Cloud Core, ApJ 1999, 526, 336, DOI 10.1086/307964.
102. [Luh04] Luhman, K. L., Peterson, D. E., Megeath, S. T., Spectroscopic Confirmation of the Least Massive Known Brown Dwarf in Chamaeleon, ApJ 2004, 617, 565, DOI 10.1086/425228.

103. [Naj00] Najita, J. R., Tiede, G. P., Carr, J. S., From Stars to Superplanets: The Low-Mass Initial Mass Function in the Young Cluster IC 348, ApJ 2000, 541, 977, DOI 10.1086/309477.
104. [Cab08a] Caballero, J. A., Dynamical parallax of σ Ori AB: mass, distance and age, MNRAS 2008, 383, 750, DOI 10.1111/j.1365-2966.2007.12614.x.
105. [Rei14] Reid, M. J., Menten, K. M., Brunthaler, A. et al., Trigonometric Parallaxes of High Mass Star Forming Regions: The Structure and Kinematics of the Milky Way, ApJ 2014, 783, 130, DOI 10.1088/0004-637X/783/2/130.
106. [Sch16] Schaefer, G. H., Hummel, C. A., Gies, D. R. et al., Orbits, Distance, and Stellar Masses of the Massive Triple Star σ Orionis, AJ 2016, 152, 213, DOI 10.3847/0004-6256/152/6/213.
107. [Kou17] Kounkel, M., Hartmann, L., Loinard, L. et al., The Gould's Belt Distances Survey (GOBELINS) II. Distances and Structure toward the Orion Molecular Clouds, ApJ 2017, 834, 142, DOI 10.3847/1538-4357/834/2/142.
108. [Cab18] Caballero, J. A., Parallactic Distances and Proper Motions of Virtually All Stars in the σ Orionis Cluster or: How I Learned to Get the Most Out of TOPCAT and Love Gaia DR2, RNAAS 2018, 2b, 25, DOI 10.3847/2515-5172/aac2b9.
109. [Bri18] Briceño, C., Calvet, N., Hernández, J. et al., The CIDA Variability Survey of Orion OB1 II: demographics of the young, low-mass stellar populations, AJ 2018, in press, eprint arXiv:1805.01008.
110. [McW97] McWilliam, A., Abundance Ratios and Galactic Chemical Evolution, ARA&A 1997, 35, 503, DOI 10.1146/annurev.astro.35.1.503.
111. [GH08] González Hernández, J. I., Caballero, J. A., Rebolo, R. et al., Chemical abundances of late-type pre-main sequence stars in the σ Orionis cluster, A&A 2008, 490, 1135, DOI 10.1051/0004-6361:200810398.
112. [Pal05] Palla, F., Randich, S., Flaccomio, E., Pallavicini, R., Age Spreads in Star-forming Regions: The Lithium Test in the Orion Nebula Cluster, ApJ 2005, 626, L49, DOI 10.1086/431668.
113. [Her07] Hernández, J., Hartmann, L., Megeath, T. et al., A Spitzer Space Telescope Study of Disks in the Young σ Orionis Cluster, ApJ 2007, 662, 1067, DOI 10.1086/513735.
114. [She08] Sherry, W. H., Walter, F. M., Wolk, S. J., Adams, N. R., Main-Sequence Fitting Distance to the σ Ori Cluster, AJ 2008, 135, 1616, DOI 10.1088/0004-6256/135/4/1616.
115. [Jef11] Jeffries, R. D., Littlefair, S. P., Naylor, T., Mayne, N. J., No wide spread of stellar ages in the Orion Nebula Cluster, MNRAS 2011, 418, 1948, DOI 10.1111/j.1365-2966.2011.19613.x.
116. [Cab08b] Caballero, J. A., Burgasser, A. J., Klement, R., Contamination by field late-M, L, and T dwarfs in deep surveys, A&A 2008, 488, 181, DOI 10.1051/0004-6361:200809520.
117. [ZO02b] Zapatero Osorio, M. R., Béjar, V. J. S., Martín, E. L., Barrado y Navascués, D., Rebolo, R., Activity at the Deuterium-burning Mass Limit in Orion, ApJ 2002, 569, L99, DOI 10.1086/340690.
118. [Luh05] Luhman, K. L., Adame, L., D'Alessio, P. et al., Discovery of a Planetary-Mass Brown Dwarf with a Circumstellar Disk, ApJ 2005, 635, L93, DOI 10.1086/498868.
119. [Cha04] Chauvin, G., Lagrange, A.-M., Dumas, C. et al., A giant planet candidate near a young brown dwarf. Direct VLT/NACO observations using IR wavefront sensing, A&A 2004, 425, L29, DOI 10.1051/0004-6361:200400056.
120. [Cha05] Chauvin, G., Lagrange, A.-M., Dumas, C. et al., Giant planet companion to 2MASSW J1207334-393254, A&A 2005, 438, L25, 10.1051/0004-6361:200500116.
121. [ZO02c] Zapatero Osorio, M. R., Béjar, V. J. S., Martín, E. L. et al., A Methane, Isolated, Planetary-Mass Object in Orion, ApJ 2002, 578, 536, DOI 10.1086/342474.
122. [MZO03] Martín, E. L., Zapatero Osorio, M. R., Spectroscopic Estimate of Surface Gravity for a Planetary Member in the σ Orionis Cluster, ApJ 2003, 593, L113, DOI 10.1086/378313.
123. [Cab06a] Caballero, J. A., Formación, evolución y multiplicidad de enanas marrones y exoplanetas gigantes, PhD thesis, Universidad de La Laguna, Spain, 2006.
124. [Cab10] Caballero, J. A., Formation, Evolution and Multiplicity of Brown Dwarfs and Giant Exoplanets, Highlights of Spanish Astrophysics V, Astrophysics and Space Science Proceedings, ISBN 978-3-642-11249-2. Springer-Verlag Berlin Heidelberg, 2010, p. 79, 10.1007/978-3-642-11250-8_7.
125. [Tru30] Trumpler, R. J., Preliminary results on the distances, dimensions and space distribution of open star clusters, Lick Observatory bulletins, 14, 154, DOI 10.5479/ADS/bib/1930LicOB.14.154T.
126. [Mer81] Mermilliod, J. C., Comparative studies of young open clusters. III - Empirical isochronous curves and the zero age main sequence, A&A 1981, 97, 235.

127. [vLe09] van Leeuwen, F., Parallaxes and proper motions for 20 open clusters as based on the new Hipparcos catalogue, A&A 2009, 497, 209, DOI 10.1051/0004-6361/200811382.
128. [Per98] Perryman, M. A. C., Brown, A. G. A., Lebreton, Y. et al., The Hyades: distance, structure, dynamics, and age, A&A 1998, 331, 81.
129. [Rei18] Reino, S., de Bruijne, J., Zari, E., d'Antona, F., Ventura, P. A Gaia study of the Hyades open cluster, MNRAS 2018, 477, 3197, DOI 10.1093/mnras/sty793.
130. [Ham93] Hambly, N. C., Hawkins, M. R. S., Jameson, R. F., Very low mass proper motion members in the Pleiades, A&AS, 1993, 100, 607.
131. [Sar14] Sarro, L. M., Bouy, H., Berihuete, A. et al., Cluster membership probabilities from proper motions and multi-wavelength photometric catalogues. I. Method and application to the Pleiades cluster, A&A 2014, 563, A45, DOI 10.1051/0004-6361/201322413.
132. [Par00] Park, B.-G., Sung, H., Bessell, M. S., Kang, Y. H., The Pre-Main-Sequence Stars and Initial Mass Function of NGC 2264, AJ 2000, 120, 894, DOI 10.1086/301459.
133. [Ven14] Venuti, L., Bouvier, J., Flaccomio, E. et al., Mapping accretion and its variability in the young open cluster NGC 2264: a study based on u-band photometry, A&A 2014, 570, A82, DOI 10.1051/0004-6361/201423776.
134. [ByN99] Barrado y Navascués, D., Stauffer, J. R., Patten, B. M., The Lithium-Depletion Boundary and the Age of the Young Open Cluster IC 2391, ApJ 1999, 522, L53, DOI 10.1086/312212.
135. [Sod14] Soderblom, D. R., Hillenbrand, L. A., Jeffries, R. D., Mamajek, E. E., Naylor, T., Ages of Young Stars, Protostars and Planets VI, 2014, Henrik Beuther, Ralf S. Klessen, Cornelis P. Dullemond, and Thomas Henning (eds.), University of Arizona Press, Tucson, p. 219, 10.2458/azu_uapress_9780816531240-ch010.
136. [Bar02] Baraffe, I., Chabrier, G., Allard, F., Hauschildt, P. H., Evolutionary models for low-mass stars and brown dwarfs: Uncertainties and limits at very young ages, A&A 2002, 382, 563, DOI 10.1051/0004-6361:20011638.
137. [Mar18] Martín, E. L., Lodieu, N., Pavlenko, Ya., Béjar, V. J. S., The Lithium Depletion Boundary and the Age of the Hyades Cluster, ApJ 2018, 856, 40, DOI 10.3847/1538-4357/aaaeb8.
138. [PR16] Peña Ramírez, K., Béjar, V. J. S., Zapatero Osorio, M. R., A new free-floating planet in the Upper Scorpius association, A&A 2016, 586, A157, DOI 10.1051/0004-6361/201527425.
139. [Lod18] Lodieu, N., Zapatero Osorio, M. R., Béjar, V. J. S., Peña Ramírez, K., The optical + infrared L dwarf spectral sequence of young planetary-mass objects in the Upper Scorpius association, MNRAS 2018, 473, 2020, DOI 10.1093/mnras/stx2279.
140. [Ber89] Bertout, C., T Tauri stars - Wild as dust, ARA&A 1989, 27, 351, DOI 10.1146/annurev.aa.27.090189.002031.
141. [Cab06b] Caballero, J. A., Martín, E. L., Zapatero Osorio, M. R. et al., S Ori J053825.4-024241: a classical T Tauri-like object at the substellar boundary, A&A 2006, 445, 143, DOI 10.1051/0004-6361:20053466.
142. [Fah13] Faherty, J. K., Rice, E. L., Cruz, K. L., Mamajek, E. E., Núñez, A., 2MASS J035523.37+113343.7: A Young, Dusty, Nearby, Isolated Brown Dwarf Resembling a Giant Exoplanet, AJ 2013, 145, 2, DOI 10.1088/0004-6256/145/1/2.
143. [Fil15] Filippazzo, J. C., Rice, E. L., Faherty, J. et al., Fundamental Parameters and Spectral Energy Distributions of Young and Field Age Objects with Masses Spanning the Stellar to Planetary Regime, ApJ 2015, 810, 158, DOI 10.1088/0004-637X/810/2/158.
144. [Fah16] Faherty, J. K., Riedel, A. R., Cruz, K. L. et al., Population Properties of Brown Dwarf Analogs to Exoplanets, ApJS 2016, 225, 10, DOI 10.3847/0067-0049/225/1/10.
145. [ZO17] Zapatero Osorio, M. R., Béjar, V. J. S., Peña Ramírez, K., Optical and Near-infrared Spectra of σ Orionis Isolated Planetary-mass Objects, ApJ 2017, 842, 65, DOI 10.3847/1538-4357/aa70ec.
146. [Bou08] Bouvier, J., Kendall, T., Meeus, G. et al., Brown dwarfs and very low mass stars in the Hyades cluster: a dynamically evolved mass function, A&A 2008, 481, 661, DOI 10.1051/0004-6361:20079303.
147. [LM08] Luhman, K. L., Muench, A. A., New Low-Mass Stars and Brown Dwarfs with Disks in the Chamaeleon I Star-Forming Region, ApJ 2008, 684, 654, DOI 10.1086/590364.
148. [Luh08a] Luhman, K. L., Allen, L. E., Allen, P. R. et al., The Disk Population of the Chamaeleon I Star-forming Region, ApJ 2008, 675, 1375, DOI 10.1086/527347.
149. [Muz15] Mužić, K., Scholz, A., Geers, V. C., Jayawardhana, R., Substellar Objects in Nearby Young Clusters (SONYC) IX: The Planetary-Mass Domain of Chamaeleon-I and Updated Mass Function in Lupus-3, ApJ 2015, 810, 159, DOI 10.1088/0004-637X/810/2/159.

150. [ByN07a] Barrado y Navascués, D., Stauffer, J. R., Morales-Calderón, M. et al., Spitzer: Accretion in Low-Mass Stars and Brown Dwarfs in the λ Orionis Cluster, ApJ 2007, 664, 481, DOI 10.1086/518816.
151. [Bay12] Bayo, A., Barrado, D., Huélamo, N. et al., Spectroscopy of very low-mass stars and brown dwarfs in the Lambda Orionis star-forming region. II. Rotation, activity and other properties of spectroscopically confirmed members of Collinder 69, A&A 2012, 547, A80, DOI 10.1051/0004-6361/201219374.
152. [Sch09] Scholz, A., Geers, V., Jayawardhana, R. et al., Substellar Objects in Nearby Young Clusters (SONYC): The Bottom of the Initial Mass Function in NGC 1333, ApJ 2009, 702, 805, DOI 10.1088/0004-637X/702/1/805.
153. [Bur09] Burgess, A. S. M., Moraux, E., Bouvier, J. et al., Young T-dwarf candidates in IC 348, A&A 2009, 508, 823, DOI 10.1051/0004-6361/200912444.
154. [Sch12] Scholz, A., Mužić, K., Geers, V. et al., Substellar Objects in Nearby Young Clusters (SONYC). IV. A Census of Very Low Mass Objects in NGC 1333, ApJ 2012, 744, 6, DOI 10.1088/0004-637X/744/1/6.
155. [AdO13] Alves de Oliveira, C., Moraux, E., Bouvier, J. et al., Spectroscopy of brown dwarf candidates in IC 348 and the determination of its substellar IMF down to planetary masses, A&A 2013, 549, A123, DOI 10.1051/0004-6361/201220229.
156. [Luh16] Luhman, K. L., Esplin, T. L., Loutrel, N. P., A Census of Young Stars and Brown Dwarfs in IC 348 and NGC 1333, ApJ 2016, 827, 52, DOI 10.3847/0004-637X/827/1/52.
157. [Esp17] Esplin, T. L., Luhman, K. L., Faherty, J. K., Mamajek, E. E., Bochanski, J. J., A Survey for Planetary-mass Brown Dwarfs in the Chamaeleon I Star-forming Region, AJ 2017, 154, 46, DOI 10.3847/1538-3881/aa74e2.
158. [Muz14] Mužić, K., Scholz, A., Geers, V. C., Jayawardhana, R., López Martí, B., Substellar Objects in Nearby Young Clusters (SONYC). VIII. Substellar Population in Lupus 3, ApJ 2014, 785, 159, DOI 10.1088/0004-637X/785/2/159.
159. [Luc06] Lucas, P. W., Weights, D. J., Roche, P. F., Riddick, F. C., Spectroscopy of planetary mass brown dwarfs in Orion, MNRAS 2006, 373, L60, DOI 10.1111/j.1745-3933.2006.00244.x.
160. [Wei09] Weights, D. J., Lucas, P. W., Roche, P. F., Pinfield, D. J., Riddick, F., Infrared spectroscopy and analysis of brown dwarf and planetary mass objects in the Orion nebula cluster, MNRAS 2009, 392, 817, DOI 10.1111/j.1365-2966.2008.14096.x.
161. [Hil13] Hillenbrand, L. A., Hoffer, A. S., Herczeg, G. J., An Enhanced Spectroscopic Census of the Orion Nebula Cluster, AJ 2013, 146, 85, DOI 10.1088/0004-6256/146/4/85.
162. [Ing14] Ingraham, P., Albert, L., Doyon, R., Artigau, E., Near-infrared (JHK) Spectroscopy of Young Stellar and Substellar Objects in Orion, ApJ 2014, 782, 8, DOI 10.1088/0004-637X/782/1/8.
163. [Sue14] Suenaga, T., Tamura, M., Kuzuhara, M. et al., Multi-object and long-slit spectroscopy of very low mass brown dwarfs in the Orion Nebular Cluster, PASJ 2014, 66, 33, DOI 10.1093/pasj/psu016.
164. [Fan16] Fang, M., Kim, J. S., Pascucci, I., Apai, D., Manara, C. F., A Candidate Planetary-mass Object with a Photoevaporating Disk in Orion, ApJ 2016, 833, L16, DOI 10.3847/2041-8213/833/2/L16.
165. [Cas07] Casewell, S. L., Dobbie, P. D., Hodgkin, S. T. et al., Proper motion L and T dwarf candidate members of the Pleiades, MNRAS 2007, 378, 1131, DOI 10.1111/j.1365-2966.2007.11848.x.
166. [Bih10] Bihain, G., Rebolo, R., Zapatero Osorio, M. R., Béjar, V. J. S., Caballero, J. A., Near-infrared low-resolution spectroscopy of Pleiades L-type brown dwarfs, A&A 2010, 519, A93, DOI 10.1051/0004-6361/200913676.
167. [Cas10] Casewell, S. L., Dobbie, P. D., Hodgkin, S. T. et al., Erratum: Proper motion L and T dwarf candidate members of the Pleiades, MNRAS 2010, 402, 1407, 10.1111/j.1365-2966.2009.16037.x.
168. [ZO14a] Zapatero Osorio, M. R., Gálvez Ortiz, M. C., Bihain, G. et al., Search for free-floating planetary-mass objects in the Pleiades, A&A 2014, 568, A77, DOI 10.1051/0004-6361/201423848.
169. [ZO14b] Zapatero Osorio, M. R., Béjar, V. J. S., Martín, E. L. et al., Spectroscopic follow-up of L- and T-type proper-motion member candidates in the Pleiades, A&A 2014, 572, A67, DOI 10.1051/0004-6361/201424634.
170. [ZO18] Zapatero Osorio, M. R., Béjar, V. J. S., Lodieu, N., Manjavacas, E., Confirming the least massive members of the Pleiades star cluster, MNRAS 2018, 475, 139, DOI 10.1093/mnras/stx3154.
171. [Hai10] Haisch, K. E., Jr., Barsony, M., Tinney, C., A Methane Imaging Survey for T Dwarf Candidates in ϱ Ophiuchi, ApJ 2010, 719, L90, DOI 10.1088/2041-8205/719/1/L90.
172. [Mar10] Marsh, K. A., Kirkpatrick, J. D., Plavchan, P., A Young Planetary-Mass Object in the ϱ Oph Cloud Core, ApJ 2010, 709, L158, DOI 10.1088/2041-8205/709/2/L158.

173. [Gee11] Geers, V., Scholz, A., Jayawardhana, R. et al., Substellar Objects in Nearby Young Clusters (SONYC). II. The Brown Dwarf Population of ϱ Ophiuchi, ApJ 2011, 726, 23, DOI 10.1088/0004-637X/726/1/23.
174. [AdO12] Alves de Oliveira, C., Moraux, E., Bouvier, J., Bouy, H., Spectroscopy of new brown dwarf members of ϱ Ophiuchi and an updated initial mass function, A&A 2012, 539, A151, DOI 10.1051/0004-6361/201118230.
175. [Muz12] Mužić, K., Scholz, A., Geers, V., Jayawardhana, R., Tamura, M., Substellar Objects in Nearby Young Clusters (SONYC). V. New Brown Dwarfs in ϱ Ophiuchi, ApJ 2012, 744, 134, DOI 10.1088/0004-637X/744/2/134.
176. [AdO13] Alves de Oliveira, C., Ábrahám, P., Marton, G. et al., Herschel survey of brown dwarf disks in ϱ Ophiuchi, A&A 2013, 559, A126, DOI 10.1051/0004-6361/201322402.
177. [CC15] Chiang, P., Chen, W. P., Discovery of Young Methane Dwarfs in the Rho Ophiuchi L 1688 Dark Cloud, ApJ 2015, 811, L16, DOI 10.1088/2041-8205/811/2/L16.
178. [Spe12] Spezzi, L., Alves de Oliveira, C., Moraux, E. et al., Searching for planetary-mass T-dwarfs in the core of Serpens, A&A 2012, 545, A105, DOI 10.1051/0004-6361/201219559.
179. [ByN01] Barrado y Navascués, D., Zapatero Osorio, M. R., Béjar, V. J. S. et al., Optical spectroscopy of isolated planetary mass objects in the σ Orionis cluster, A&A 2001, 377, L9, DOI 10.1051/0004-6361:20011152.
180. [Mar01] Martín, E. L., Zapatero Osorio, M. R., Barrado y Navascués, D., Béjar, V. J. S., Rebolo, R., Keck NIRC Observations of Planetary-Mass Candidate Members in the σ Orionis Open Cluster, ApJ 2011, 558, L117, DOI 10.1086/323633.
181. [Cab04] Caballero, J. A., Béjar, V. J. S., Rebolo, R., Zapatero Osorio, M. R., Photometric variability of young brown dwarfs in the σ Orionis open cluster, A&A 2004, 424, 857, DOI 10.1051/0004-6361:20047048.
182. [ZO07] Zapatero Osorio, M. R., Caballero, J. A., Béjar, V. J. S. et al., Discs of planetary-mass objects in σ Orionis, A&A 2007, 472, L9, DOI 10.1051/0004-6361:20078116.
183. [Bih09] Bihain, G., Rebolo, R., Zapatero Osorio, M. R. et al., Candidate free-floating super-Jupiters in the young σ Orionis open cluster, A&A 2009, 506, 1169, DOI 10.1051/0004-6361/200912210.
184. [Lod09] Lodieu, N., Zapatero Osorio, M. R., Rebolo, R., Martín, E. L., Hambly, N. C., A census of very-low-mass stars and brown dwarfs in the σ Orionis cluster, A&A 2009, 505, 1115, DOI 10.1051/0004-6361/200911966.
185. [Béj11] Béjar, V. J. S., Zapatero Osorio, M. R., Rebolo, R. et al., The Substellar Population of σ Orionis: A Deep Wide Survey, ApJ 2011, 743, 64, DOI 10.1088/0004-637X/743/1/64.
186. [PR12] Peña Ramírez, K., Béjar, V. J. S., Zapatero Osorio, M. R., Petr-Gotzens, M. G., Martín, E. L., New Isolated Planetary-mass Objects and the Stellar and Substellar Mass Function of the σ Orionis Cluster, ApJ 2012, 754, 30, DOI 10.1088/0004-637X/754/1/30.
187. [PR15] Peña Ramírez, K., Zapatero Osorio, M. R., Béjar, V. J. S., Characterization of the known T-type dwarfs towards the σ Orionis cluster, A&A 2015, 574, A118, DOI 10.1051/0004-6361/201424816.
188. [Luh09a] Luhman, K. L., Mamajek, E. E., Allen, P. R., Cruz, K. L., An Infrared/X-Ray Survey for New Members of the Taurus Star-Forming Region, ApJ 2009, 703, 399, DOI 10.1088/0004-637X/703/1/399.
189. [EL17] Esplin, T. L., Luhman, K. L., A Survey For Planetary-mass Brown Dwarfs in the Taurus and Perseus Star-forming Regions, AJ 2017, 154, 134, DOI 10.3847/1538-3881/aa859b.
190. [Lod07] Lodieu, N., Hambly, N. C., Jameson, R. F. et al., New brown dwarfs in Upper Sco using UKIDSS Galactic Cluster Survey science verification data, MNRAS 2007, 374, 372, DOI 10.1111/j.1365-2966.2006.11151.x.
191. [Lod08] Lodieu, N., Hambly, N. C., Jameson, R. F., Hodgkin, S. T., Near-infrared cross-dispersed spectroscopy of brown dwarf candidates in the UpperSco association, MNRAS 2008, 383, 1385, DOI 10.1111/j.1365-2966.2007.12676.x.
192. [Lod11] Lodieu, N., Hambly, N. C., Dobbie, P. D. et al., Testing the fragmentation limit in the Upper Sco association, MNRAS 2011, 418, 2604, DOI 10.1111/j.1365-2966.2011.19651.x.
193. [Lod13] Lodieu, N., Dobbie, P. D., Cross, N. J. G. et al., Probing the Upper Scorpius mass function in the planetary-mass regime, MNRAS 2013, 435, 2474, DOI 10.1093/mnras/stt1460.
194. [BM17] Béjar, V. J. S., Martín, E. L., Brown dwarfs and free-floating planets in young stellar clusters, In Handbook of exoplanets; Deeg, H., Belmonte, J. A., Eds.; Springer, Cham, 2017. DOI 10.1007/978-3-319-30648-3_92-1.

195. [Bur04b] Burgasser, A. J., Kirkpatrick, J. D., McGovern, M. R. et al., S Orionis 70: Just a Foreground Field Brown Dwarf?, ApJ 2004, 604, 827, DOI 10.1086/382129.
196. [Luh08] Luhman, K. L., Hernández, J., Downes, J. J., Hartmann, L., Briceño, C., Disks around Brown Dwarfs in the σ Orionis Cluster, ApJ 2008, 688, 362, DOI 10.1086/592264.
197. [SJ08] Scholz, A., Jayawardhana, R., Dusty Disks at the Bottom of the Initial Mass Function, ApJ 2008, 672, L49, DOI 10.1086/526340.
198. [ZO08] Zapatero Osorio, M. R., Béjar, V. J. S., Bihain, G. et al., New constraints on the membership of the T dwarf S Ori 70 in the σ Orionis cluster, A&A 2008, 477, 895, DOI 10.1051/0004-6361:20078600.
199. [Eli78] Elias, J. H., A study of the Taurus dark cloud complex, ApJ 1978, 224, 857, DOI 10.1086/156436.
200. [KH95] Kenyon, S. J., Hartmann, L., Pre-Main-Sequence Evolution in the Taurus-Auriga Molecular Cloud, ApJS 1995, 101, 117, DOI 10.1086/192235.
201. [And05] Andrews, S. M., Williams, J. P., Circumstellar Dust Disks in Taurus-Auriga: The Submillimeter Perspective, ApJ 2005, 631, 1134, DOI 10.1086/432712.
202. [Wal94] Walter, F. M., Vrba, F. J., Mathieu, R. D., Brown, A., Myers, P. C., X-ray sources in regions of star formation. 5: The low mass stars of the Upper Scorpius association, AJ 1999, 107, 692, DOI 10.1086/116889.
203. [Pre02] Preibisch, T., Brown, A. G. A., Bridges, T., Guenther, E., Zinnecker, H., Exploring the Full Stellar Population of the Upper Scorpius OB Association, AJ 2002, 124, 404, DOI 10.1086/341174.
204. [Pec12] Pecaut, M. J., Mamajek, E. E., Bubar, E. J., A Revised Age for Upper Scorpius and the Star Formation History among the F-type Members of the Scorpius-Centaurus OB Association, ApJ 2012, 746, 154, DOI 10.1088/0004-637X/746/2/154.
205. [Her47] Hertzsprung, E., Catalogue de 3259 étoiles dans les Pléiades, Annalen van de Sterrewacht te Leiden, 1947, 19, A1.
206. [Sod93] Soderblom, D. R., Jones, B. F., Balachandran, S. et al., The evolution of the lithium abundances of solar-type stars. III - The Pleiades, AJ 1993, 106, 1059, DOI 10.1086/116704.
207. [Mel14] Melis, C., Reid, M. J., Mioduszewski, A. J., Stauffer, J. R., Bower, G. C., A VLBI resolution of the Pleiades distance controversy, Science 2014, 345, 1029, DOI 10.1126/science.1256101.
208. [She04] Sherry, W. H., Walter, F. M., Wolk, S. J., Photometric Identification of the Low-Mass Population of Orion OB1b. I. The σ Orionis Cluster, AJ 2004, 128, 2316, DOI 10.1086/424863.
209. [SD15] Simón-Díaz, S., Caballero, J. A., Lorenzo, J. et al., Orbital and Physical Properties of the σ Ori Aa, Ab, B Triple System, ApJ 2015, 799, 169, DOI 10.1088/0004-637X/799/2/169.
210. [Lee68] Lee, T. A., Interstellar extinction in the Orion association, ApJ 1968, 152, 913, DOI 10.1086/149607.
211. [Cow79] Cowie, L. L., Songaila, A., York, D. G., Orion's Cloak - A rapidly expanding shell of gas centered on the Orion OB1 association, ApJ 1979, 230, 469, DOI 10.1086/157103.
212. [DM01] Dolan, C. J., Mathieu, R. D., The Spatial Distribution of the λ Orionis Pre-Main-Sequence Population, AJ 2001, 121, 2124, DOI 10.1086/319946.
213. [ByN04] Barrado y Navascués, D., Stauffer, J. R., Bouvier, J., Jayawardhana, R., Cuillandre, J.-C., The Substellar Population of the Young Cluster λ Orionis, ApJ 2004, 610, 1064, DOI 10.1086/421762.
214. [BH15] Bowler, B. P., Hillenbrand, L. A., Near-infrared Spectroscopy of 2M0441+2301 AabBab: A Quadruple System Spanning the Stellar to Planetary Mass Regimes, ApJ 2015, 811, L30, DOI 10.1088/2041-8205/811/2/L30.
215. [Cac15] Cáceres, C., Hardy, A., Schreiber, M. R. et al., On the Nature of the Tertiary Companion to FW Tau: ALMA CO Observations and SED Modeling, ApJ 2015, 806, L22, DOI 10.1088/2041-8205/806/2/L22.
216. [Luh09b] Luhman, K. L., Mamajek, E. E., Allen, P. R., Muench, A. A., Finkbeiner, D. P., Discovery of a Wide Binary Brown Dwarf Born in Isolation, ApJ 2009, 691, 1265, DOI 10.1088/0004-637X/691/2/1265.
217. [Bes17a] Best, W. M. J., Liu, M. C., Magnier, E. A. et al., A Search for L/T Transition Dwarfs with Pan-STARRS1 and WISE. III. Young L Dwarf Discoveries and Proper Motion Catalogs in Taurus and Scorpius-Centaurus, ApJ 2017, 837, 95, DOI 10.3847/1538-4357/aa5df0.
218. [Dob05] Dobashi, K., Uehara, H., Kandori, R. et al., Atlas and Catalog of Dark Clouds Based on Digitized Sky Survey I, PASJ 2005, 57, S1, DOI 10.1093/pasj/57.sp1.S1.
219. [Yor00] York, D. G., Adelman, J., Anderson, J. E., Jr. et al., The Sloan Digital Sky Survey: Technical Summary, AJ 2000, 120, 1579, DOI 10.1086/301513.
220. [Faz04] Fazio, G. G., Hora, J. L., Allen, L. E. et al., The Infrared Array Camera (IRAC) for the Spitzer Space Telescope, ApJS 2004, 154, 10, DOI 10.1086/422843.

221. [Skr06] Skrutskie, M. F., Cutri, R. M., Stiening, R. et al., The Two Micron All Sky Survey (2MASS), AJ 2006, 131, 1163, DOI 10.1086/498708.
222. [Law07] Lawrence, A., Warren, S. J., Almaini, O. et al., The UKIRT Infrared Deep Sky Survey (UKIDSS), MNRAS 2007, 379, 1599, DOI 10.1111/j.1365-2966.2007.12040.x.
223. [Kai10] Kaiser, N., Burgett, W., Chambers, K. et al., The Pan-STARRS wide-field optical/NIR imaging survey, SPIE 2010, 7733, E0E, DOI 10.1117/12.859188.
224. [Wri10] Wright, E. L., Eisenhardt, P. R. M., Mainzer, A. K. et al., The Wide-field Infrared Survey Explorer (WISE): Mission Description and Initial On-orbit Performance, AJ 2010, 140, 1868, DOI 10.1088/0004-6256/140/6/1868.
225. [Gai16] Gaia Collaboration, Brown, A. G. A., Vallenari, A., Prusti, T. et al., Gaia Data Release 1. Summary of the astrometric, photometric, and survey properties, A&A 2016, 595, A2, DOI 10.1051/0004-6361/201629512.
226. [Luh17] Luhman, K. L., Mamajek, E. E., Shukla, S. J., Loutrel, N. P., A Survey for New Members of the Taurus Star-forming Region with the Sloan Digital Sky Survey, AJ 2017, 153, 46, DOI 10.3847/1538-3881/153/1/46.
227. [Kra17] Kraus, A. L., Herczeg, G. J., Rizzuto, A. C. et al., The Greater Taurus-Auriga Ecosystem. I. There is a Distributed Older Population, ApJ 2017, 838, 150, DOI 10.3847/1538-4357/aa62a0.
228. [Cab08c] Caballero, J. A., Spatial distribution of stars and brown dwarfs in σ Orionis, MNRAS 2008, 383, 375, 10.1111/j.1365-2966.2007.12555.x.
229. [Jef06] Jeffries, R. D., Maxted, P. F. L., Oliveira, J. M., Naylor, T., Kinematic structure in the young σ Orionis association, MNRAS 2006, 371, L6, DOI 10.1111/j.1745-3933.2006.00196.x.
230. [Cab08d] Caballero, J. A., Stars and brown dwarfs in the σ Orionis cluster: the Mayrit catalogue, A&A 2008, 478, 667, 10.1051/0004-6361:20077885.
231. [Laf08] Lafrenière, D., Jayawardhana, R., van Kerkwijk, M. H., Direct Imaging and Spectroscopy of a Planetary-Mass Candidate Companion to a Young Solar Analog, ApJ 2008, 689, L153, DOI 10.1086/595870.
232. [Laf10] Lafrenière, D., Jayawardhana, R., van Kerkwijk, M. H., The Directly Imaged Planet Around the Young Solar Analog 1RXS J160929.1-210524: Confirmation of Common Proper Motion, Temperature, and Mass, ApJ 2010, 719, 497, DOI 10.1088/0004-637X/719/1/497.
233. [Ire11] Ireland, M. J., Kraus, A., Martinache, F., Law, N., Hillenbrand, L. A., Two Wide Planetary-mass Companions to Solar-type Stars in Upper Scorpius, ApJ 2011, 726, 113, DOI 10.1088/0004-637X/726/2/113.
234. [Lac15] Lachapelle, F.-R., Lafrenière, D., Gagné, J. et al., Characterization of Low-mass, Wide-separation Substellar Companions to Stars in Upper Scorpius: Near-infrared Photometry and Spectroscopy, ApJ 2015, 802, 61, DOI 10.1088/0004-637X/802/1/61.
235. [Laf11] Lafrenière, D., Jayawardhana, R., Janson, M. et al., Discovery of an ~23 $M_{Jup}$ Brown Dwarf Orbiting ~700 AU from the Massive Star HIP 78530 in Upper Scorpius, ApJ 2011, 730, 42, DOI 10.1088/0004-637X/730/1/42.
236. [Bor10] Borucki, W. J., Koch, D., Basri, G. et al., Kepler Planet-Detection Mission: Introduction and First Results, Science 2010, 327, 977, DOI 10.1126/science.1185402.
237. [Lis14] Lissauer, J. J., Marcy, G. W., Bryson, S. T. et al., Validation of Kepler's Multiple Planet Candidates. II. Refined Statistical Framework and Descriptions of Systems of Special Interest, ApJ 2014, 784, 44, DOI 10.1088/0004-637X/784/1/44.
238. [Dav16] David, T. J., Hillenbrand, L. A., Petigura, E. A. et al., A Neptune-sized transiting planet closely orbiting a 5-10-million-year-old star, Nature 2016, 534, 658, DOI 10.1038/nature18293.
239. [Man16] Mann, A. W., Newton, E. R., Rizzuto, A. C. et al., Zodiacal Exoplanets in Time (ZEIT). III. A Short-period Planet Orbiting a Pre-main-sequence Star in the Upper Scorpius OB Association, AJ 2016, 152, 61, DOI 10.3847/0004-6256/152/3/61.
240. [Van16] Vanderburg, A., Latham, D. W., Buchhave, L. A. et al., Planetary Candidates from the First Year of the K2 Mission, ApJS 2016, 222, 14, DOI 10.3847/0067-0049/222/1/14.
241. [Mar98] Martín, E. L., Basri, G., Zapatero Osorio, M. R., Rebolo, R., García López, R. J., The First L-Type Brown Dwarf in the Pleiades, ApJ 1998, 507, L41, DOI 10.1086/311675.
242. [Bih06] Bihain, G., Rebolo, R., Béjar, V. J. S. et al., Pleiades low-mass brown dwarfs: the cluster L dwarf sequence, A&A 2006, 458, 805, DOI 10.1051/0004-6361:20065124.

243. [Lod12] Lodieu, N., Deacon, N. R., Hambly, N. C., Astrometric and photometric initial mass functions from the UKIDSS Galactic Clusters Survey - I. The Pleiades, MNRAS 2012, 422, 1495, DOI 10.1111/j.1365-2966.2012.20723.x.
244. [Cas11] Casewell, S. L., Jameson, R. F., Burleigh, M. R. et al., Methane band and Spitzer mid-IR imaging of L and T dwarf candidates in the Pleiades, MNRAS 2011, 412, 2071, DOI 10.1111/j.1365-2966.2010.18044.x.
245. [Gau15] Gauza, B., Béjar, V. J. S., Pérez-Garrido, A. et al., Discovery of a Young Planetary Mass Companion to the Nearby M Dwarf VHS J125601.92-125723.9, ApJ 2015, 804, 96, DOI 10.1088/0004-637X/804/2/96.
246. [Jam02] Jameson, R. F., Dobbie, P. D., Hodgkin, S. T., Pinfield, D. J., Brown dwarfs in the Pleiades: spatial distribution and mass function, MNRAS 2002, 335, 853, DOI 10.1046/j.1365-8711.2002.05667.x.
247. [Mam15] Mamajek, E. E., A Moving Cluster Distance to the Exoplanet 2M1207b in the TW Hydrae Association, ApJ 2005, 634, 1385, DOI 10.1086/468181.
248. [Son06] Song, I., Schneider, G., Zuckerman, B. et al., HST NICMOS Imaging of the Planetary-mass Companion to the Young Brown Dwarf 2MASSW J1207334-393254, ApJ 2006, 652, 724, DOI 10.1086/507831.
249. [Duc08] Ducourant, C., Teixeira, R., Chauvin, G. et al., An accurate distance to 2M1207Ab, A&A 2008, 477, L1, DOI 10.1051/0004-6361:20078886.
250. [Pat10] Patience, J., King, R. R., de Rosa, R. J., Marois, C., The highest resolution near infrared spectrum of the imaged planetary mass companion 2M1207 b, A&A 2010, 517, A76, DOI 10.1051/0004-6361/201014173.
251. [Bar11] Barman, T. S., Macintosh, B., Konopacky, Q. M., Marois, C., The Young Planet-mass Object 2M1207b: A Cool, Cloudy, and Methane-poor Atmosphere, ApJ 2011, 735, L39, DOI 10.1088/2041-8205/735/2/L39.
252. [Zho16] Zhou, Y., Apai, D., Schneider, G. H., Marley, M. S., Showman, A. P., Discovery of Rotational Modulations in the Planetary-mass Companion 2M1207b: Intermediate Rotation Period and Heterogeneous Clouds in a Low Gravity Atmosphere, ApJ 2016, 818, 176, DOI 10.3847/0004-637X/818/2/176.
253. [BC07] Biller, B. A., Close, L. M., A Direct Distance and Luminosity Determination for a Self-luminous Giant Exoplanet: The Trigonometric Parallax to 2MASSW J1207334-393254Ab, ApJ 2007, 669, L41, DOI 10.1086/523799.
254. [Moh07] Mohanty, S., Jayawardhana, R., Huélamo, N., Mamajek, E. E., The Planetary Mass Companion 2MASS 1207-3932B: Temperature, Mass, and Evidence for an Edge-on Disk, ApJ 2007, 657, 1064, DOI 10.1086/510877.
255. [MM07] Mamajek, E. E., Meyer, M. R., An Improbable Solution to the Underluminosity of 2M1207B: A Hot Protoplanet Collision Afterglow, ApJ, 2007, 668, L175, DOI 10.1086/522957.
256. [Ske11] Skemer, A. J., Close, L. M., Szűcs, L. et al., Evidence Against an Edge-on Disk Around the Extrasolar Planet, 2MASS 1207 b and a New Thick-cloud Explanation for Its Underluminosity, ApJ 2011, 732, 107, DOI 10.1088/0004-637X/732/2/107.
257. [Ria12] Riaz, B., Lodato, G., Stamatellos, D., Gizis, J. E., Herschel SPIRE observations of the TWA brown dwarf disc 2MASSW J1207334-393254, MNRAS 2012, 422, L6, DOI 10.1111/j.1745-3933.2012.01225.x.
258. [Ric17] Ricci, L., Cazzoletti, P., Czekala, I. et al., ALMA Observations of the Young Substellar Binary System 2M1207, AJ 2017, 154, 24, DOI 10.3847/1538-3881/aa78a0.
259. [Cab06c] Caballero, J. A., Martín, E. L., Dobbie, P. D., Barrado y Navascués, D., Are isolated planetary-mass objects really isolated?. A brown dwarf-exoplanet system candidate in the σ Orionis cluster, A&A 2006, 460, 635, DOI 10.1051/0004-6361:20066162.
260. [SE04] Scholz, A., Eislöffel, J., Rotation and accretion of very low mass objects in the σ Ori cluster, A&A 2004, 419, 249, DOI 10.1051/0004-6361:20034022.
261. [Fra06] Franciosini, E., Pallavicini, R., Sanz-Forcada, J., XMM-Newton observations of the σ Orionis cluster. II. Spatial and spectral analysis of the full EPIC field, A&A 2006, 446, 501, DOI 10.1051/0004-6361:20053605.
262. [ByN07b] Barrado y Navascués, D., Bayo, A., Morales-Calderón, M. et al., The young, wide and very low mass visual binary Lambda Orionis 167, A&A 2007, 468, L5, DOI 10.1051/0004-6361:20077258.
263. [Béj08] Béjar, V. J. S., Zapatero Osorio, M. R., Pérez-Garrido, A. et al., Discovery of a Wide Companion near the Deuterium-burning Mass Limit in the Upper Scorpius Association, ApJ 2008, 673, L185, DOI 10.1086/527557.
264. [Tod10] Todorov, K. O., Luhman, K. L., McLeod, K. K., Discovery of a Planetary-mass Companion to a Brown Dwarf in Taurus, ApJ 2010, 714, L84, DOI 10.1088/2041-8205/714/1/L84.

265. [Tod14] Todorov, K. O., Luhman, K. L., Konopacky, Q. M. et al., A Search for Companions to Brown Dwarfs in the Taurus and Chamaeleon Star-Forming Regions, ApJ 2014, 788, 40, DOI 10.1088/0004-637X/788/1/40.
266. [JI06] Jayawardhana, R., Ivanov, V. D., Discovery of a Young Planetary-Mass Binary, Science 2006, 313, 1279, DOI 10.1126/science.1132128.
267. [Bra06] Brandeker, A., Jayawardhana, R., Ivanov, V. D., Kurtev, R, Infrared Spectroscopy of the Ultra-Low-Mass Binary Oph 162225-240515, ApJ 2006, 653, L61, DOI 10.1086/510308.
268. [Clo07] Close, L. M., Zuckerman, B., Song, I. et al., The Wide Brown Dwarf Binary Oph 1622-2405 and Discovery of a Wide, Low-Mass Binary in Ophiuchus (Oph 1623-2402): A New Class of Young Evaporating Wide Binaries?, ApJ 2007, 660, 1492, DOI 10.1086/513417.
269. [Luh07] Luhman, K. L., Allers, K. N., Jaffe, D. T. et al., Ophiuchus 1622-2405: Not a Planetary-Mass Binary, ApJ 2007, 659, 1629, DOI 10.1086/512539.
270. [Cab09] Caballero, J. A., Reaching the boundary between stellar kinematic groups and very wide binaries. The Washington double stars with the widest angular separations, A&A 2009, 507, 251, DOI 10.1051/0004-6361/200912596.
271. [Lod05] Lodato, G., Delgado-Donate, E., Clarke, C. J., Constraints on the formation mechanism of the planetary mass companion of 2MASS 1207334-393254, MNRAS 2005, 364, L91, DOI 10.1111/j.1745-3933.2005.00112.x.
272. [Des99] Desidera, S., Properties of Hypothetical Planetary Systems around the Brown Dwarf Gliese 229B, PASP 1999, 111, 1529, DOI 10.1086/316467.
273. [CR02] Caballero, J. A., Rebolo, R., Variability in brown dwarfs: atmospheres and transits, In: Proceedings of the First Eddington Workshop on Stellar Structure and Habitable Planet Finding, 11 - 15 June 2001, Córdoba, Spain. Editor: B. Battrick, Scientific editors: F. Favata, I. W. Roxburgh & D. Galadi. ESA SP-485, Noordwijk: ESA Publications Division, ISBN 92-9092-781-X, 2002, p. 261.
274. [Del08] Delorme, P., Delfosse, X., Albert, L. et al., CFBDS J005910.90-011401.3: reaching the T-Y brown dwarf transition?, A&A 2008, 482, 961, DOI 10.1051/0004-6361:20079317.
275. [Cus11] Cushing, M. C., Kirkpatrick, J. D., Gelino, C. R. et al., The Discovery of Y Dwarfs using Data from the Wide-field Infrared Survey Explorer (WISE), ApJ 2011, 743, 50, DOI 10.1088/0004-637X/743/1/50.
276. [Kir11] Kirkpatrick, J. D., Cushing, M. C., Gelino, C. R. et al., The First Hundred Brown Dwarfs Discovered by the Wide-field Infrared Survey Explorer (WISE), ApJS 2011, 197, 19, DOI 10.1088/0067-0049/197/2/19.
277. [Kir12] Kirkpatrick, J. D., Gelino, C. R., Cushing, M. C. et al., Further Defining Spectral Type "Y" and Exploring the Low-mass End of the Field Brown Dwarf Mass Function, ApJ 2012, 753, 156, DOI 10.1088/0004-637X/753/2/156.
278. [Liu12] Liu, M. C., Dupuy, T. J., Bowler, B. P., Leggett, S. K., Best, W. M. J., Two Extraordinary Substellar Binaries at the T/Y Transition and the Y-band Fluxes of the Coolest Brown Dwarfs, ApJ 2012, 758, 57, DOI 10.1088/0004-637X/758/1/57.
279. [Tin12] Tinney, C. G., Faherty, J. K., Kirkpatrick, J. D. et al., WISE J163940.83-684738.6: A Y Dwarf Identified by Methane Imaging, ApJ 2012, 759, 60, DOI 10.1088/0004-637X/759/1/60.
280. [Bei13] Beichman, C., Gelino, C. R., Kirkpatrick, J. D. et al., The Coldest Brown Dwarf (or Free-floating Planet)?: The Y Dwarf WISE 1828+2650, ApJ 2013, 764, 101, DOI 10.1088/0004-637X/764/1/101.
281. [Kir13] Kirkpatrick, J. D., Cushing, M. C., Gelino, C. R. et al., Discovery of the Y1 Dwarf WISE J064723.23-623235.5, ApJ 2013, 776, 128, DOI 10.1088/0004-637X/776/2/128.
282. [DK13] Dupuy, T. J., Kraus, A. L., Distances, Luminosities, and Temperatures of the Coldest Known Substellar Objects, Science 2013, 341, 1492, DOI 10.1126/science.1241917
283. [Cus14] Cushing, M. C., Kirkpatrick, J. D., Gelino, C. R., Three New Cool Brown Dwarfs Discovered with the Wide-field Infrared Survey Explorer (WISE) and an Improved Spectrum of the Y0 Dwarf WISE J041022.71+150248.4, AJ 2014, 147, 113, DOI 10.1088/0004-6256/147/5/113.
284. [Pin14] Pinfield, D. J., Gromadzki, M., Leggett, S. K. et al., Discovery of a new Y dwarf: WISE J030449.03-270508.3, MNRAS 2014, 444, 1931, DOI 10.1093/mnras/stu1540.
285. [Liu13] Liu, M. C., Magnier, E. A., Deacon, N. R. et al., The Extremely Red, Young L Dwarf PSO J318.5338-22.8603: A Free-floating Planetary-mass Analog to Directly Imaged Young Gas-giant Planet, ApJ 2013, 777, L20, DOI 10.1088/2041-8205/777/2/L20.
286. [Mac13] Mace, G. N., Kirkpatrick, J. D., Cushing, M. C. et al., A Study of the Diverse T Dwarf Population Revealed by WISE, ApJS 2013, 205, 6, DOI 10.1088/0067-0049/205/1/6.

287. [Gag14a] Gagné, J., Lafrenière, D., Doyon, R., Malo, L., Artigau, É., BANYAN. II. Very Low Mass and Substellar Candidate Members to Nearby, Young Kinematic Groups with Previously Known Signs of Youth, ApJ 2014, 783, 121, DOI 10.1088/0004-637X/783/2/121.
288. [Gag14b] Gagné, J., Faherty, J., K., Cruz, K. et al., The Coolest Isolated Brown Dwarf Candidate Member of TWA, ApJ 2014, 785, L14, DOI 10.1088/2041-8205/785/1/L14.
289. [Gag15a] Gagné, J., Burgasser, A. J., Faherty, J. K. et al., SDSS J111010.01+011613.1: A New Planetary-mass T Dwarf Member of the AB Doradus Moving Group, ApJ 2015, 808, L20, DOI 10.1088/2041-8205/808/1/L20.
290. [Gag15b] Gagné, J., Faherty, J. K., Cruz, K. L. et al., BANYAN. VII. A New Population of Young Substellar Candidate Members of Nearby Moving Groups from the BASS Survey, ApJS 2015, 219, 33, DOI 10.1088/0067-0049/219/2/33.
291. [Kel15] Kellogg, K., Metchev, S., Geißler, K. et al., A Targeted Search for Peculiarly Red L and T Dwarfs in SDSS, 2MASS, and WISE: Discovery of a Possible L7 Member of the TW Hydrae Association, AJ 2015, 150, 182, DOI 10.1088/0004-6256/150/6/182.
292. [Sch16] Schneider, A. C., Windsor, J., Cushing, M. C., Kirkpatrick, J. D., Wright, E. L., WISEA J114724.10-204021.3: A Free-floating Planetary Mass Member of the TW Hya Association, ApJ 2016, 822, L1, DOI 10.3847/2041-8205/822/1/L1.
293. [Bes17b] Best, W. M. J., Liu, M. C., Dupuy, T. J., Magnier, E. A., The Young L Dwarf 2MASS J11193254-1137466 Is a Planetary-mass Binary, ApJ 2017, 843, L4, DOI 10.3847/2041-8213/aa76df.
294. [Gag17a] Gagné, J., Faherty, J. K., Mamajek, E. E. et al., BANYAN. IX. The Initial Mass Function and Planetary-mass Object Space Density of the TW HYA Association, ApJS 2017, 228, 18, DOI 10.3847/1538-4365/228/2/18.
295. [Sch18b] Schneider, A. C., Hardegree-Ullman, K. K., Cushing, M. C., Kirkpatrick, J. D., Shkolnik, E. L., Spitzer Light Curves of the Young, Planetary-mass TW Hya Members 2MASS J11193254–1137466AB and WISEA J114724.10–204021.3, AJ 2018, 155, 238, DOI 10.3847/1538-3881/aabfc2.
296. [Gag17b] Gagné, J., Faherty, J. K., Burgasser, A. J. et al., SIMP J013656.5+093347 Is Likely a Planetary-mass Object in the Carina-Near Moving Group, ApJ 2017, 841, L1, DOI 10.3847/2041-8213/aa70e2.
297. [Gag18] Gagné, J., Allers, K. N., Theissen, C. A. et al., 2MASS J13243553+6358281 Is an Early T-type Planetary-mass Object in the AB Doradus Moving Group, ApJ 2018, 854, L27, DOI 10.3847/2041-8213/aaacfd.
298. [Del12] Delorme, P., Gagné, J., Malo, L. et al., CFBDSIR2149-0403: a 4-7 Jupiter-mass free-floating planet in the young moving group AB Doradus?, A&A 2012, 548, A26, DOI 10.1051/0004-6361/201219984.
299. [Del17] Delorme, P., Dupuy, T., Gagné, J. et al., CFBDSIR 2149-0403: young isolated planetary-mass object or high-metallicity low-mass brown dwarf?, A&A 2017, 602, A82, DOI 10.1051/0004-6361/201629633.
300. [Ske16] Skemer, A. J., Morley, C. V., Allers, K. N. et al., The First Spectrum of the Coldest Brown Dwarf, ApJ 2016, 826, L17, DOI 10.3847/2041-8205/826/2/L17.
301. [ZO16] Zapatero Osorio, M. R., Lodieu, N., Béjar, V. J. S. et al., Near-infrared photometry of WISE J085510.74-071442.5, A&A 2016, 592, A80, DOI 10.1051/0004-6361/201628662.
302. [Luh11] Luhman, K. L., Burgasser, A. J., Bochanski, J. J., Discovery of a Candidate for the Coolest Known Brown Dwarf, ApJ 2011, 730, L9, DOI 10.1088/2041-8205/730/1/L9.
303. [Rod11] Rodríguez, D. R., Zuckerman, B., Melis, C., Song, I., The Ultra Cool Brown Dwarf Companion of WD 0806-661B: Age, Mass, and Formation Mechanism, ApJ 2011, 732, L29, DOI 10.1088/2041-8205/732/2/L29.
304. [Luh12] Luhman, K. L., Burgasser, A. J., Labbé, I. et al., Confirmation of One of the Coldest Known Brown Dwarfs, ApJ 2012, 744, 135, DOI 10.1088/0004-637X/744/2/135.
305. [Luh14b] Luhman, K. L., Morley, C. V., Burgasser, A. J., Esplin, T. L., Bochanski, J. J., Near-infrared Detection of WD 0806-661 B with the Hubble Space Telescope, ApJ 2014, 794, 16, DOI 10.1088/0004-637X/794/1/16.
306. [Kop14] Kopytova, T. G., Crossfield, I. J. M., Deacon, N. R. et al., Deep z-band Observations of the Coolest Y Dwarf, ApJ 2014, 797, 3, DOI 10.1088/0004-637X/797/1/3.
307. [LE14] Luhman, K. L., Esplin, T. L., A New Parallax Measurement for the Coldest Known Brown Dwarf, ApJ 2014, 796, 6, DOI 10.1088/0004-637X/796/1/6.
308. [Tin14] Tinney, C. G., Faherty, J. K., Kirkpatrick, J. D. et al., The Luminosities of the Coldest Brown Dwarfs, ApJ 2014, 796, 39, DOI 10.1088/0004-637X/796/1/39.

309. [Wri14] Wright, E. L., Mainzer, A., Kirkpatrick, J. D. et al., NEOWISE-R Observation of the Coolest Known Brown Dwarf, AJ 2014, 148, 82, DOI 10.1088/0004-6256/148/5/82.
310. [Yat17] Yates, J. S., Palmer, P. I., Biller, B., Cockell, C. S., Atmospheric Habitable Zones in Y Dwarf Atmospheres, ApJ 2017, 836, 184, DOI 10.3847/1538-4357/836/2/184.
311. [Kum64] Kumar, S. S., On the nature of red stars of low luminosity, The Observatory 1964, 84, 18.
312. [Kum67] Kumar, S. S., On planets and black dwarfs, Icarus 1967, 6, 136, 10.1016/0019-1035(67)90012-7.
313. [LLB76] Low, C., Lynden-Bell, D., The minimum Jeans mass or when fragmentation must stop, MNRAS 1976, 176, 367, DOI 10.1093/mnras/176.2.367.
314. [Ree76] Rees, M. J., Opacity-limited hierarchical fragmentation and the masses of protostars, MNRAS 1976, 176, 483, DOI 10.1093/mnras/176.3.483.
315. [Sil77] Silk, J., On the fragmentation of cosmic gas clouds. II - Opacity-limited star formation, ApJ 1977, 214, 152, DOI 10.1086/155240.
316. [Toh80] Tohline, J. E., The gravitational fragmentation of primordial gas clouds, ApJ 1980, 239, 417, DOI 10.1086/158125.
317. [Bos89] Boss, A. P., Low-mass star and planet formation, PASP 1989, 101, 767, DOI 10.1086/132499.
318. [Bos01] Boss, A. P., Formation of Planetary-Mass Objects by Protostellar Collapse and Fragmentation, ApJ 2001, 551, L167, DOI 10.1086/320033.
319. [Bat03] Bate, M. R., Bonnell, I. A., Bromm, V., The formation of a star cluster: predicting the properties of stars and brown dwarfs, MNRAS 2003, 339, 577, DOI 10.1046/j.1365-8711.2003.06210.x.
320. [Whi07] Whitworth, A., Bate, M. R., Nordlund, Å., Reipurth, B., Zinnecker, H., The Formation of Brown Dwarfs: Theory, Protostars and Planets V, B. Reipurth, D. Jewitt, and K. Keil (eds.), University of Arizona Press, Tucson, 2007, p.459.
321. [Luh07] Luhman, K. L., Joergens, V., Lada, C., Muzerolle, J., Pascucci, I., White, R., The Formation of Brown Dwarfs: Observations, Protostars and Planets V, B. Reipurth, D. Jewitt, and K. Keil (eds.), University of Arizona Press, Tucson, 2007, p.443.
322. [Bos00] Boss, A. P., Possible Rapid Gas Giant Planet Formation in the Solar Nebula and Other Protoplanetary Disks, ApJ 2000, 536, L101, DOI 10.1086/312737.
323. [RC01] Reipurth, B., Clarke, C., The Formation of Brown Dwarfs as Ejected Stellar Embryos, AJ 2001, 122, 432, DOI 10.1086/321121.
324. [Bat02] Bate, M. R., Bonnell, I. A., Bromm, V., The formation of close binary systems by dynamical interactions and orbital decay, MNRAS 2002, 336, 705, DOI 10.1046/j.1365-8711.2002.05775.x.
325. [MB98] Marcy, G. W., Butler, R. P., Detection of Extrasolar Giant Planets, ARA&A 1998, 36, 57, 10.1146/annurev.astro.36.1.57.
326. [Par14] Parker, R. J., Wright, N. J., Goodwin, S. P., Meyer, M. R., Dynamical evolution of star-forming regions, MNRAS 2014, 438, 620, DOI 10.1093/mnras/stt2231.
327. [Cab07b] Caballero, J. A., A near-infrared/optical/X-ray survey in the centre of σ Orionis, AN 2007, 328, 917, DOI 10.1002/asna.200710778.
328. [Bou09] Bouy, H., Huélamo, N., Martín, E. L. et al., A deep look into the cores of young clusters. I. σ Orionis, A&A 2009, 493, 931, DOI 10.1051/0004-6361:200810267.
329. [Pal12] Palau, A., de Gregorio-Monsalvo, I., Morata, Ó. et al., A search for pre-substellar cores and proto-brown dwarf candidates in Taurus: multiwavelength analysis in the B213-L1495 clouds, MNRAS 2012, 424, 2778, DOI 10.1111/j.1365-2966.2012.21390.x.
330. [Gah13] Gahm, G. F., Persson, C. M., Mäkelä, M. M., Haikala, L. K., Mass and motion of globulettes in the Rosette Nebula, A&A 2013, 555, A57, DOI 10.1051/0004-6361/201321547.
331. [Pal14] Palau, A., Zapata, L. A., Rodríguez, L. F. et al., IC 348-SMM2E: a Class 0 proto-brown dwarf candidate forming as a scaled-down version of low-mass stars, MNRAS 2014, 444, 833, DOI 10.1093/mnras/stu1461.
332. [Haw15] Haworth, T. J., Facchini, S., Clarke, C. J., The theory of globulettes: candidate precursors of brown dwarfs and free-floating planets in H II regions, MNRAS 2015, 446, 1098, DOI 10.1093/mnras/stu2174.
333. [Mor15] Morata, Ó., Palau, A., González, R. F. et al., First Detection of Thermal Radiojets in a Sample of Proto-brown Dwarf Candidates, ApJ 2015, 807, 55, DOI 10.1088/0004-637X/807/1/55.
334. [dGM16] de Gregorio-Monsalvo, I., Barrado, D., Bouy, H. et al., A submillimetre search for pre- and proto-brown dwarfs in Chamaeleon II, A&A 2016, 590, A79, DOI 10.1051/0004-6361/201424149.

335. [Liu16] Liu, T., Zhang, Q., Kim, K.-T. et al., Planck Cold Clumps in the λ Orionis Complex. I. Discovery of an Extremely Young Class 0 Protostellar Object and a Proto-brown Dwarf Candidate in the Bright-rimmed Clump PGCC G192.32-11.88, ApJS 2016, 222, 7, DOI 10.3847/0067-0049/222/1/7.
336. [Bay17] Bayo, A., Joergens, V., Liu, Y. et al., First Millimeter Detection of the Disk around a Young, Isolated, Planetary-mass Object, ApJ 2017, 841, L11, DOI 10.3847/2041-8213/aa7046.
337. [Ria17] Riaz, B., Briceño, C., Whelan, E. T., Heathcote, S., First Large-scale Herbig-Haro Jet Driven by a Proto-brown Dwarf, ApJ 2017, 844, 47, DOI 10.3847/1538-4357/aa70e8.
338. [LSS09] LSST Science Collaboration, Abell, P. A., Allison, J., Anderson, S. F. et al., LSST Science Book, Version 2.0, LSST Corporation, 2009, eprint arXiv:0912.0201.
339. [Lau11] Laureijs, R., Amiaux, J., Arduini, S. et al., Euclid Definition Study Report, ESA 2011, eprint arXiv:1110.3193.
340. [Spe15] Spergel, D., Gehrels, N., Baltay, C. et al., Wide-Field InfrarRed Survey Telescope-Astrophysics Focused Telescope Assets WFIRST-AFTA 2015 Report, NASA 2015, eprint arXiv:1503.03757.
341. [Dea18] Deacon, N. R., Detecting free-floating planets using water-depend colour terms in the next generation of infrared space-based surveys, MNRAS 2018, in press, eprint arXiv:1808.04828.
342. [Bur03b] Burrows, A., Sudarsky, D., Lunine, J. I., Beyond the T Dwarfs: Theoretical Spectra, Colors, and Detectability of the Coolest Brown Dwarfs, ApJ 2003, 596, 587, DOI 10.1086/377709.
343. [Tre17] Tremblin, P., Chabrier, G., Baraffe, I. et al., Cloudless Atmospheres for Young Low-gravity Substellar Objects, ApJ 2017, 850, 46, DOI 10.3847/1538-4357/aa9214.
344. [Mor18] Morley, C. V., Skemer, A. J., Allers, K. N. et al., An L Band Spectrum of the Coldest Brown Dwarf, ApJ 2018, 858, 97, DOI 10.3847/1538-4357/aabe8b.
345. [GS07] Gilmozzi, R., Spyromilio, J., The European Extremely Large Telescope (E-ELT), Messenger 2007, 127, 11.
346. [bot] A Dictionary of Color Terms in Botanical Latin. Avaliable online: http://www.applet-magic.com/botlatin.htm (accessed on 22 Aug 2018).
347. [Hip18] Hippke, M., Spaceflight from Super-Earths is difficult, International Journal of Astrobiology 2018, in press, eprint arXiv:1804.04727.
348. [How18] Howe, A. R., A Tether-Assisted Space Launch System for Super-Earths, astro-ph 2018, eprint arXiv:1805.06438.
349. [Fou12] Foullon, C., Playing Music with Magnetic Fields, Astronomical Society of the Pacific Conference Series 2002, 259, 482, ISBN 1-58381-099-4.
350. [Fra06] Fraknoi, A. The Music of the Spheres in Education: Using Astronomically Inspired Music, Astronomy Education Review 2006, 5a, 139, DOI 10.3847/AER2006009.
351. [Ula09] Ulaş, B., Musical scale estimation for some multiperiodic pulsating stars, Communications in Asteroseismology, 159, 131, DOI 10.1553/cia159s131.
352. [Cab10] Caballero, J. A., González Sánchez, S., Caballero, I., Music and Astronomy, Astrophysics and Space Science Proceedings 2010, 548, DOI 10.1007/978-3-642-11250-8_170.
353. [Lub10] Lubowich, D., Music and Astronomy Under the Stars 2009, Astronomical Society of the Pacific Conference Series 2010, 431, 47.
354. [Fra12] Fraknoi, A., Music Inspired by Astronomy: A Resource Guide Organized by Topic, Astronomy Education Review 2012, 11, 010303, DOI 10.3847/AER2012043.
355. [Bat12] Bate, M. R., Stellar, brown dwarf and multiple star properties from a radiation hydrodynamical simulation of star cluster formation, MNRAS 2012, 419, 3115, DOI 10.1111/j.1365-2966.2011.19955.x.
356. [mul] Videoclip musical de "El ordenador simula el nacimiento de las estrellas". Available online: https://www.youtube.com/watch?v=J9lCSCV3Mkk (accessed on 22 Aug 2018).
357. [Cab17] Caballero, J. A., Arias, A., Machuca, J. J., Morente, S., Music and astronomy. II. Unitedsoundofcosmos, Highlights on Spanish Astrophysics IX, Proceedings of the XII Scientific Meeting of the Spanish Astronomical Society held on July 18-22, 2016, in Bilbao, Spain, ISBN 978-84-617-8931-3. S. Arribas, A. Alonso-Herrero, F. Figueras, C. Hernández-Monteagudo, A. Sánchez-Lavega, S. Pérez-Hoyos (eds.), 2017, p. 715.
358. [Mar03] Martín, E. L., Brown Dwarfs, Proceedings of IAU Symposium #211, held 20-24 May 2002 at University of Hawaii, Honolulu, Hawaii. Edited by Eduardo Martín. San Francisco: Astronomical Society of the Pacific, 2003.

359. [Bos03] Boss, A. P., Basri, G., Kumar, S. S. et al., Nomenclature: Brown Dwarfs, Gas Giant Planets, and ?, Brown Dwarfs, Proceedings of IAU Symposium #211, held 20-24 May 2002 at University of Hawaii, Honolulu, Hawaii. Edited by Eduardo Martín. San Francisco: Astronomical Society of the Pacific, 2003, p. 529.